\documentclass[onecolumn,11pt,preprintnumbers,superscriptaddress,amsmath,amssymb]{revtex4}

\usepackage{graphicx}% Include figure files
\usepackage{dcolumn}% Align table columns on decimal point
\usepackage{bm}% bold math
\usepackage{theorem}
\usepackage{color}
\usepackage{multirow}
\usepackage{mathrsfs}
\usepackage{mathtools}

{\theorembodyfont{\normalfont\it}
\theoremheaderfont{\normalfont\bf}
\newtheorem{thm}{ Theorem}
\newtheorem{dfn}[thm]{ Definition}
\newtheorem{lmm}[thm]{ Lemma}

\newtheorem{crl}[thm]{ Corollary}
\newtheorem{asm}[thm]{ Assumption}
\newtheorem{prp}[thm]{ Proposition}
\newtheorem{cjt}[thm]{ Conjecture}
\newtheorem{rmk}[thm]{ Remark}}

{\theorembodyfont{\normalfont}
\theoremheaderfont{\normalfont\it}
\newtheorem{prf}{ Proof:}}

{\theorembodyfont{\normalfont}
\theoremheaderfont{\normalfont\it}
}

{\theorembodyfont{\normalfont}
\theoremheaderfont{\normalfont\it}
}

{\theorembodyfont{\normalfont}
\theoremheaderfont{\normalfont\it}
}

{\theorembodyfont{\normalfont}
\theoremheaderfont{\normalfont\it}
}

{\theorembodyfont{\normalfont}
\theoremheaderfont{\normalfont\it}
}

\newcommand{\bra}[1]{\mbox{$\langle#1|$}}

\newcommand{\ket}[1]{\mbox{$|#1\rangle$}}

\newcommand{\proj}[1]{\mbox{$\ket{#1}\!\bra{#1}$}}

\newcommand{\alg}[1]{\begin{align}#1\end{align}}

\newcommand{\nn}{\nonumber}
\newcommand{\ca}[1]{{\mathcal #1}}
\newcommand{\mbb}[1]{{\mathbb #1}}
\newcommand{\mfk}[1]{{\mathfrak #1}}

\newcommand{\bthm}[1]{\begin{thm}\label{thm:#1}}
\newcommand{\ethm}{\end{thm}}
\newcommand{\rthm}[1]{\ref{thm:#1}}
\newcommand{\rThm}[1]{Theorem \ref{thm:#1}}
\newcommand{\blmm}[1]{\begin{lmm}\label{lmm:#1}}
\newcommand{\elmm}{\end{lmm}}
\newcommand{\rLmm}[1]{Lemma \ref{lmm:#1}}

\newcommand{\bdfn}[1]{\begin{dfn}\label{dfn:#1}}
\newcommand{\edfn}{\end{dfn}}

\newcommand{\basm}[1]{\begin{asm}\label{asm:#1}}
\newcommand{\easm}{\end{asm}}

\newcommand{\bprp}[1]{\begin{prp}\label{prp:#1}}
\newcommand{\eprp}{\end{prp}}

\newcommand{\bcrl}[1]{\begin{crl}\label{crl:#1}}
\newcommand{\ecrl}{\end{crl}}

\newcommand{\bcjt}[1]{\begin{cjt}\label{cjt:#1}}
\newcommand{\ecjt}{\end{cjt}}

\newcommand{\brmk}[1]{\begin{rmk}\label{rmk:#1}}
\newcommand{\ermk}{\end{rmk}}

\newcommand{\bprf}{\begin{prf}}
\newcommand{\eprf}{\end{prf}}
\newcommand{\laeq}[1]{\label{eq:#1}}
\newcommand{\req}[1]{(\ref{eq:#1})}

\newcommand{\QED}{\hfill$\blacksquare$}
\newcommand{\lsec}[1]{\label{sec:#1}}

\newcommand{\rSec}[1]{Section \ref{sec:#1}}
\newcommand{\lapp}[1]{\label{app:#1}}

\newcommand{\rApp}[1]{Appendix \ref{app:#1}}

\newcommand{\bitem}{\begin{itemize}}
\newcommand{\entem}{\end{itemize}}
\newcommand{\benum}{\begin{enumerate}}
\newcommand{\ennum}{\end{enumerate}}

\newcommand{\otm}{\otimes}

\newcommand{\rFig}[1]{Figure \ref{fig:#1}}

\newcommand{\argmax}{\mathop{\rm arg~max}\limits}
\newcommand{\argmin}{\mathop{\rm arg~min}\limits}

\begin{document}

%\preprint{APS/123-QED}

\title{Exact Exponent for Atypicality of Random Quantum States}% Force line breaks with \\

\author{Eyuri Wakakuwa}
\email{e.wakakuwa@gmail.com}
\affiliation{Department of Mathematical Informatics, Graduate School of Informatics,
Nagoya University, 
Furo-cho, Chikusa-ku, Nagoya, 464-8601, 
Japan}%

\date{\today}% It is always \today, today,

\begin{abstract}
We study the properties of the random quantum states induced from the uniformly random pure states on a bipartite quantum system by taking the partial trace over the larger subsystem.
Most of the previous studies have adopted a viewpoint of ``concentration of measure'' and have focused on the behavior of the states close to the average.
In contrast, we investigate the large deviation regime, where the states may be far from the average.
We prove the following results:
First, the probability that the induced random state is within a given set decreases no slower or faster than exponential in the dimension of the subsystem traced out.
Second, the exponent is equal to the quantum relative entropy of the maximally mixed state and the given set, multiplied by the dimension of the remaining subsystem.
Third, the total probability of a given set strongly concentrates around the element closest to the maximally mixed state, a property that we call {\it conditional concentration}.
Along the same line, we also investigate an asymptotic behavior of coherence of random pure states in a single system with large dimensions.
\end{abstract}

%\pacs{03.67.Bg, 03.67.Mn}% PACS, the Physics and Astronomy
                       
                                                      
\maketitle

\section{Introduction}

The concept of {\it concentration of measure} has played significant roles in quantum information theory and the foundation of quantum statistical mechanics.
It asserts that functions on high-dimensional spaces are almost constant if it is sufficiently continuous (see e.g.~\cite{ledoux2001concentration,milman2009asymptotic}).
In the context of quantum information theory, Ref.~\cite{hayden2006aspects} applied this concept to analyzing entanglement properties of multipartite quantum states.
Ref.~\cite{gross2009most} proved that almost all pure states are useless for universal quantum computation.
Ref.~\cite{aubrun2011hastings} provided a simple derivation of the counterexample \cite{hastings2009superadditivity} to the additivity conjecture of the classical capacity of quantum channels. 
As for the foundation of quantum statistical mechanics, Ref.~\cite{popescu2006entanglement} applied the concept of concentration of measure to prove that a physical system would be in the thermal state for almost all pure states of the universe.
Ref.~\cite{reimann2015generalization} generalized von Neumann's approach to prove thermalization of isolated quantum many-body systems, among many others.

Most of the previous researches along these lines are based on Levy’s lemma.
Ref.~\cite{hayden2006aspects} proved that any function defined on the set of the pure states takes values close to the average for the overwhelming majority of the pure states if the dimension of the Hilbert space is sufficiently large and the function is Lipschitz continuous. 
Mathematically, the concentration inequalities obtained from Levy's lemma are of the form
\alg{
{\rm Pr}_\Psi\{|f(\Psi)-\mbb{E}_\Psi f(\Psi)|>\epsilon\}
\leq
2\exp{\left(-\frac{\kappa(d)\epsilon^2}{\eta^2}\right)}.
\laeq{kanna}
}
Here, $\Psi$ is a pure state that is randomly chosen from the Hilbert space according to the unitary invariant measure, $f$ is a function defined on the set of the unit vectors therein, $\eta$ is the Lipschitz constant of $f$, $d$ is the dimension of the Hilbert space, and $\kappa(d)$ is a positive linear function of $d$.
The inequalities of the type \req{kanna} apply to any finite $d$ and any Lipschitz function.
Because of this generality, Levy's lemma has been widely applied to prove the concentration of functions of random quantum states.
However, the approaches from Levy's lemma have limitations in that (i)~they provide us only {\it upper} bounds on the probability of the value of $f$ distant from the average, and (ii) they apply only to the properties that are represented by the deviation of Lipschitz functions.

In this paper, we address the above problem by following the spirit of the large deviation theory~\cite{dembo2009large}.
Our primal focus is on the random quantum states obtained from the uniformly random pure states in a bipartite system by taking the partial trace over the larger subsystem. 
This scenario was originally investigated in \cite{lloyd1988complexity,zyczkowski2001induced}.
To be concrete, we consider the random state ${\rm Tr}_E[|\Psi\rangle\!\langle\Psi|]$, where $|\Psi\rangle$ is the pure state on a composite system $SE$ that obeys the unitary invariant measure.
We focus on an asymptotic limit of the dimension of $E$ going to infinity, while the dimension of $S$ is assumed to be fixed. 
Our goal is to evaluate the probability of the state ${\rm Tr}_E[|\Psi\rangle\!\langle\Psi|]$ within a given set $\Omega$. 

The results are as follows: 
First, we prove that the probability of $\Omega$ converges to zero, no slower than or no faster than, exponential in the dimension of $E$ unless $\Omega$ contains the maximally mixed state.  
Second, we prove that the exponent is equal to the relative entropy of the maximally mixed state $\pi$ against $\Omega$, multiplied by the dimension of $S$.
Namely, we prove the equation
\alg{
-\lim_{d_E\rightarrow\infty}\frac{1}{d_E}\ln{\rm Pr}_\Psi\{{\rm Tr}_E[\proj{\Psi}^{SE}]\in\Omega\}=d_S\inf_{\rho\in\Omega}D(\pi\|\rho),
\laeq{maiz}
} 
where $\Psi$ is the uniformly random pure quantum state on $SE$.
In the proof, we exploit the fact that the unitary invariant measure is represented by the multivariate normal distribution (see e.g.~Section 7.2 in \cite{watrous2018theory}), to which we apply Sanov’s theorem for continuous random variables in the version of \cite{groeneboom1979large}. 
Third, we prove a property that we call {\it conditional concentration}: 
When we already know that the state is within a set $\Omega$, the probability that it is not close to the optimal state (in terms of the relative entropy from $\pi$) is exponentially small. 
Hence, the total probability of $\Omega$ strongly concentrates around the element of $\Omega$ closest to the maximally mixed state. 
Finally, we show that a similar method is applicable to analyzing the large deviation property of coherence of random pure states in a {\it single} quantum system.

The main result, Eq.~\req{maiz}, is applicable to any subset of states that is regular in a topological sense. 
Hence, it is possible to obtain a large deviation bound in the case where the property of the states of our interest is not necessarily represented by the deviation of Lipschitz functions.
Moreover, Eq.~\req{maiz} provides not only an upper bound but also a lower bound on the probability of the subset in the asymptotic limit.
That is, Eq.~\req{maiz} is both a concentration bound and an anti-concentration bound.
It is because of this novelty that we can prove the property of conditional concentration. 
A drawback of Eq.~\req{maiz} from Eq.~\req{kanna} is that it is valid only in the asymptotic limit of the subsystem dimension to infinity.
It is, therefore, left as an open problem to obtain a large deviation bound that applies to finite $d$ and is tight in the asymptotic limit.

This paper is organized as follows.
In \rSec{formresults}, we present the formulation of the problem and state the main results.
\rSec{Sanovcont} reviews a version of Sanov's theorem in general probability spaces and applies it to prove a variant of Sanov's theorem for continuous random variables.
The proofs of the main results are presented in \rSec{LDP}.
In \rSec{examples}, we apply the main results to several cases where the exponents can be calculated explicitly.
\rSec{condconc} introduces the concept of conditional concentration and provides a proof thereof.
In \rSec{coherence}, we investigate the large deviation property of quantum coherence of random pure states in a single Hilbert space.
Conclusions are given in \rSec{conclusion}.

\subsection*{Notations}
%When $M$  and $N$ are linear operators on ${\mathcal H}^A$ and ${\mathcal H}^B$, respectively, we denote $M\otimes N$ as $M^A\otimes N^B$ for clarity. 
The sets of the linear operators, the positive semidefinite operators and the normalized density operators on a Hilbert space $\ca{H}$ are denoted by $\ca{L}(\ca{H})$, $\ca{P}(\ca{H})$ and $\ca{S}(\ca{H})$, respectively.
The identity operator on $\ca{H}$ is denoted by $I$.  
$\ln{x}$ represents the natural logarithm of $x$.
For $p\geq1$, the Schatten $p$-norm of $W\in\ca{L}(\ca{H})$ is defined by $\|W\|_p:=({\rm Tr}[(W^\dagger W)^{p/2}])^{1/p}$ (see e.g.~Section 1.1.3 in \cite{watrous2018theory}).
For any set $S$ in a topological space or a metric space, its closure, interior and boundary are denoted by ${\rm cl}S$, ${\rm int}S$ and $\partial S$, respectively.

For quantum states $\sigma,\rho\in\ca{S}(\ca{H})$, the quantum relative entropy is defined by
\alg{
D(\sigma\|\rho)
:=
{\rm Tr}[\sigma\ln{\sigma}-\sigma\ln{\rho}].
}
For a finite alphabet $\ca{X}$ and probability distributions $P\equiv\{p(x)\}_{x\in\ca{X}}$ and $Q\equiv\{q(x)\}_{x\in\ca{X}}$, the classical relative entropy is defined by
\alg{
D(P\|Q)
:=
\sum_{x\in\ca{X}}p(x)\ln{\frac{p(x)}{q(x)}}.
}
For binary probability distributions $\{\alpha,1-\alpha\}$ and $\{\beta,1-\beta\}$, we define the binary relative entropy $\mathscr{D}$ by 
\alg{
\mathscr{D}(\alpha\|\beta)
&
:=D(\{\alpha,1-\alpha\}\|\{\beta,1-\beta\}).
%\\
%&=
%\alpha\ln{\frac{\alpha}{\beta}}+(1-\alpha)\ln{\frac{1-\alpha}{1-\beta}}.
\laeq{binaryRE}
} 
For the properties of the quantum and classical relative entropies, see e.g.~\cite{wildetext,cover05}.

\section{Formulation and Main Result}
\lsec{formresults}

Let $\ca{H}^S$ and $\ca{H}^E$ be finite-dimensional Hilbert spaces such that
\alg{
\dim{\ca{H}^S}=m,
\;
\dim{\ca{H}^E}=n.
}
Let the Hilbert space of the whole system be
\alg{
\ca{H}=\ca{H}^S\otm\ca{H}^E.
}
For each $\ket{\Psi}\in\ca{H}$, we define
\alg{
\rho_\Psi:={\rm Tr}_E[\proj{\Psi}].
}
We consider the unitarily invariant measure on $\ca{H}$.
To clarify the dependence on the dimension of $\ca{H}^E$, we denote it by $\mu_n$.
When $|\Psi\rangle$ is randomly chosen according to $\mu_n$, we represent it as $\Psi\sim\mu_n$.
The induced random state $\rho_\Psi\in\ca{S}(\ca{H}^S)$ is referred to as the {\it rank-$n$ random state} \cite{hayden2006aspects}.
We are interested in the probability of the rank-$n$ random state $\rho_\Psi$ within a given subset $\Omega\subseteq\ca{S}(\ca{H}^S)$, namely ${\rm Pr}\left\{\left.\rho_\Psi\in\Omega\right|\Psi\sim\mu_n\right\}$.
We focus on the asymptotic behavior of the probability in the limit of $n$ to infinity.
We regard $\ca{S}(\ca{H}^S)$ as a metric space equipped with the trace distance.
The following theorem is the main result of this paper:

\bthm{LDP}%{\bf[Large Deviation.]}
For any subset $\Omega\subseteq\ca{S}(\ca{H}^S)$, it holds that
\alg{
-\limsup_{n\rightarrow\infty}\frac{1}{n}\ln{{\rm Pr}\left\{\left.\rho_\Psi\in\Omega\right|\Psi\sim\mu_n\right\}}
&\geq
m\inf_{\rho\in{\rm cl}\Omega}D(\pi\|\rho),
\laeq{gojupe}
\\
-\liminf_{n\rightarrow\infty}\frac{1}{n}\ln{{\rm Pr}\left\{\left.\rho_\Psi\in\Omega\right|\Psi\sim\mu_n\right\}}
&\leq
m\inf_{\rho\in{\rm int}\Omega}D(\pi\|\rho),
\laeq{gojure}
}
where $\pi\equiv I/m$ is the maximally mixed state on $S$. 
In particular, for any subset $\Omega\subseteq\ca{S}(\ca{H}^S)$ satisfying ${\rm cl}({\rm int}\Omega)={\rm cl}\Omega$,
it holds that
\alg{
-\lim_{n\rightarrow\infty}\frac{1}{n}\ln{{\rm Pr}\left\{\left.\rho_\Psi\in\Omega\right|\Psi\sim\mu_n\right\}}
=
m\inf_{\rho\in\Omega}D(\pi\|\rho).
\laeq{reibou}
}
\ethm

\noindent
In the terminology of the large deviation theory \cite{dembo2009large}, \rThm{LDP} asserts that the probability density of $\rho_\Psi$ satisfies the large deviation principle with the rate function $mD(\pi\|\rho)$.
A proof of \rThm{LDP} will be provided in \rSec{LDP}, based on the preliminary results presented in \rSec{Sanovcont}.

\begin{figure}[t]
\begin{center}
\includegraphics[bb={0 30 423 423}, scale=0.4]{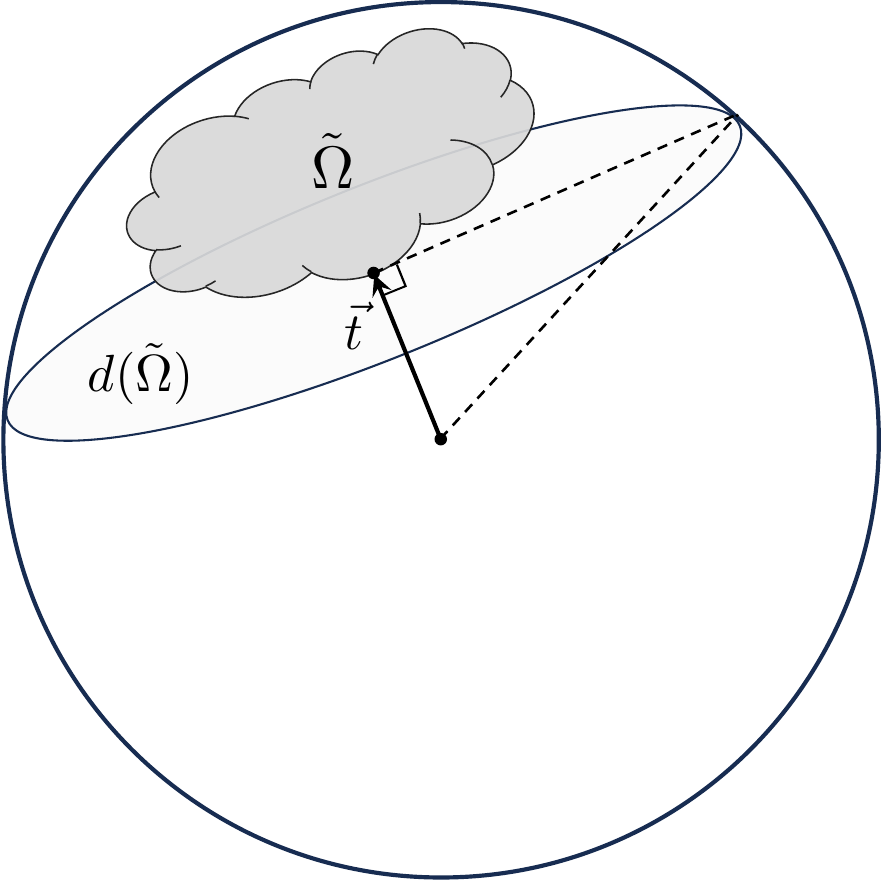}
\end{center}
\caption{The example of a one-qubit system is depicted. 
Every given subset $\Omega\subset\ca{S}(\mbb{C}^2)$ is represented by a subset $\tilde{\Omega}$ of the Bloch ball.
Let $\vec{t}$ be the Bloch vector of the state that is closest to the maximally mixed state $\pi$. 
Let $S$ be the plane that is tangent to $\tilde{\Omega}$ at point $\vec{t}$.
Let $d(\tilde{\Omega})$ be the section of the Bloch sphere by $S$.
The log of the inverse radius of $d(\tilde{\Omega})$ is equal to the relative entropy $D(\pi\|\rho(\vec{t}))$.
The twice of it is the exponent for $\Omega$.
}
\label{fig:bloch}
\end{figure}

We present a few examples for which the exponent can be calculated concretely.
More applications will be provided in \rSec{examples}.\\

{\bf Example 1 (qubit system):}
We consider a one-qubit system where $m=2$.
Every state $\rho$ is represented as
\alg{
\rho=\rho(\vec{t}):=\frac{1}{2}I+\frac{1}{2}\sum_{\kappa=x,y,z}t_\kappa\sigma_\kappa,
\laeq{dotour}
}
where $\vec{t}\equiv(t_x,t_y,t_z)$ is a three dimensional real vector satisfying $|\vec{t}|:=\sqrt{t_1^2+t_2^2+t_3^2}\leq1$ and $(\sigma_x,\sigma_y,\sigma_z)$ is the Pauli operators defined by
\alg{
\sigma_x=
\begin{pmatrix}
0&1\\
1&0
\end{pmatrix},\:
\sigma_y=
\begin{pmatrix}
0&-i\\
i&0
\end{pmatrix},\:
\sigma_z=
\begin{pmatrix}
1&0\\
0&-1
\end{pmatrix}
}
(see e.g.~\cite{nielsentext}).
The eigenvalues of $\rho(\vec{t})$ are $(1\pm|\vec{t}|)/2$.
Thus, we have
\alg{
D(\pi\|\rho(\vec{t}))
=
\ln{\left(\frac{1}{\sqrt{1-|\vec{t}|^2}}\right)}.
\laeq{ft1-|}
}
The exponent is equal to zero if and only if $\vec{t}=\vec{0}$ (i.e., $\rho(\vec{t})=\pi$), and goes to infinity when $\vec{t}$ approaches the surface of the Bloch sphere (i.e., $|\vec{t}|\rightarrow1$).
Every subset $\Omega\subset\ca{S}(\mbb{C}^2)$ is mapped by the correspondence \req{dotour} to a subset $\tilde{\Omega}$ of the unit ball in $\mbb{R}^3$.
Thus, we have
\alg{
 \inf_{\rho\in\Omega}D(\pi\|\rho)
=
\inf_{\vec{t}\in\tilde{\Omega}}D(\pi\|\rho(\vec{t}))
=
\inf_{\vec{t}\in\tilde{\Omega}}
\ln {\left( \frac{1}{\sqrt{1 - |\vec{t}|^2}} \right)},
}
which is depicted in \rFig{bloch}.
\QED\\

\begin{figure}[t]
\begin{center}
\includegraphics[bb={10 20 263 190}, scale=1.0]{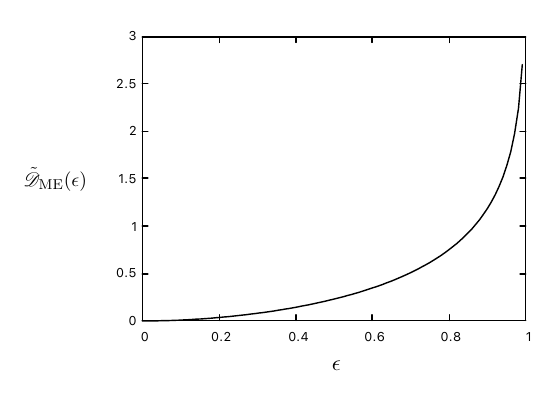}
\end{center}
\caption{
A graph of the function $\tilde{\mathscr{D}}_{\rm ME}(\epsilon):=\mathscr{D}_{\rm ME}(\epsilon;m=3)$, which is given by \req{tsuson}, is presented.
It goes to infinity in the limit of $\epsilon\rightarrow1$.
}
\label{fig:MEV}
\end{figure}

{\bf Example 2 (Maximum Eigenvalue):}
Let $\lambda_{\rm max}(\rho)$ denote the maximum eigenvalue of $\rho\in\ca{S}(\ca{H}^S)$.
We consider the set of states such that the maximum eigenvalue is no smaller than or equal to $(1+(m-1)\epsilon)/m$, where $0<\epsilon<1$.
The infimum of the relative entropy is evaluated as
\alg{
\inf_{\rho:\lambda_{\rm max}(\rho)\geq \frac{1+(m-1)\epsilon}{m}}D(\pi\|\rho)
=\inf_{\rho:\lambda_{\rm max}(\rho)=\frac{1+(m-1)\epsilon}{m}}D(\pi\|\rho)
=\mathscr{D}_{\rm ME}(\epsilon;m),
}
where
\alg{
\mathscr{D}_{\rm ME}(\epsilon;m)
:=
-\frac{1}{m}
\left[\ln{\left(1+(m-1)\epsilon\right)}+(m-1)\ln{\left(1-\epsilon\right)}\right].
\laeq{tsuson}
}
Indeed, the infimum is achieved by a state
\alg{
\tilde{\rho}_\epsilon=
\frac{1+(m-1)\epsilon}{m}\proj{1}
+
\frac{1-\epsilon}{m}\sum_{k=2}^m\proj{k}.
}
Note that every state satisfying $\lambda_{\rm max}(\rho)\geq(1+(m-1)\epsilon)/m$ can be transformed to $\varrho$ by a random unitary operation, under which $D(\pi\|\rho)$ is non-increasing.
A graph of the function $\mathscr{D}_{\rm ME}(\epsilon;m)$ for the case of $m=3$ is presented in \rFig{MEV}.
In the second order of $\epsilon$, the function \req{tsuson} is equal to $(m-1)\epsilon^2/2$.
This is a constant factor improvement over the large deviation bound obtained from Lemma III.4 in \cite{hayden2006aspects} (see \rApp{compPRPR} for the detail).
\QED\\

\begin{figure}[t]
\begin{center}
\includegraphics[bb={10 20 288 198}, scale=0.95]{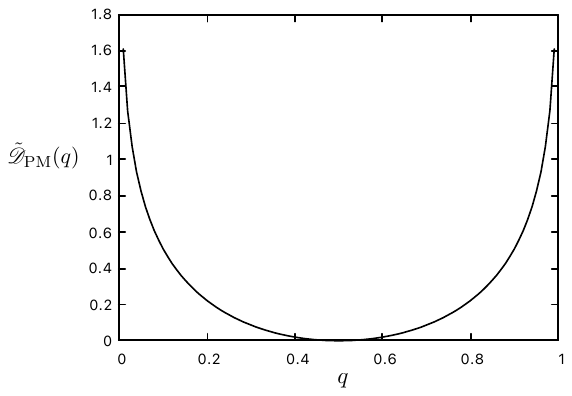}
\end{center}
\caption{
A graph of the function $\tilde{\mathscr{D}}_{\rm PM}(q):=\mathscr{D}(m_0/m=1/2\|q)$ is presented.
It goes to infinity in the limit of $q\rightarrow0,1$.
}
\label{fig:binaryRE}
\end{figure}

{\bf Example 3 (Binary Projective Measurement):}
Consider a binary projective measurement on $S$, which is represented by a pair $\{\Pi_0,\Pi_1\}$ of orthogonal projectors on $\ca{H}^S$ such that $\Pi_0+\Pi_1=I$.
The probability of obtaining an outcome $s=0,1$ when the above measurement is performed on a state $\rho$ is described by the probability distribution $P_\rho\equiv\{p_\rho,1-p_\rho\}$, where $p_\rho={\rm Tr}[\Pi_0\rho]$.
Let $Q\equiv\{q,1-q\}$ be a binary probability distribution.
The variational distance (see e.g.~\cite{cover05}) between $P_\rho$ and $Q$ is calculated to be $|p_\rho-q|$.
We consider the set of states such that $|p_\rho-q|$ is no greater than $\delta$.
We have that
\alg{
\inf_{\rho:|p_\rho-q|\leq\delta}D(\pi\|\rho)
=
\inf_{P':|p'-q|\leq\delta}\inf_{\rho:P_\rho=P'}D(\pi\|\rho),
}
where $P'\equiv\{p',1-p'\}$.
The monotonicity of the relative entropy implies
\alg{
&\inf_{\rho:P_\rho=P'}D(\pi\|\rho)
\geq
\inf_{\rho:P_\rho=P'}D(P_\pi\|P_\rho)
=
\mathscr{D}\left(\left. \frac{m_0}{m}\right\|p'\right),
}
where $m_0:={\rm rank}\Pi_0$ and $\mathscr{D}$ is the binary relative entropy defined by \req{binaryRE}.
The equality holds for the state $\rho_{P'}:=p'\pi_0+(1-p')\pi_1$, where $\pi_s:=\Pi_s/{\rm Tr}[\Pi_s]$ for $s=0,1$.
Hence,
\alg{
\inf_{\rho:|p_\rho-q|\leq\delta}D(\pi\|\rho)
=
\inf_{P':|p'-q|\leq\delta}\mathscr{D}\left(\left. \frac{m_0}{m}\right\|p'\right).
}
Due to the continuity of the binary relative entropy, we obtain
\alg{
\lim_{\delta\rightarrow0}\inf_{\rho:|p_\rho-q|\leq\delta}D(\pi\|\rho)
=
\mathscr{D}\left(\left. \frac{m_0}{m}\right\|q\right).
}
A graph of this function for the case of $m_0/m=1/2$ is depicted in \rFig{binaryRE}.
\QED\\

We will use the following properties of the quantum relative entropy in the subsequent sections:

\blmm{Drhoprop}
The relative entropy $D(\pi\|\rho)$ satisfies the following properties:
\begin{enumerate}
\item \label{DLB} Lower Bound: $D(\pi\|\rho)\geq0$ with the equality if and only if $\rho=\pi$.
\item \label{DUB} Upper Bound: $D(\pi\|\rho)<\infty$ if and only if $\rho$ is a full-rank state.
\item \label{DCv}  Convexity: For any $\rho,\sigma\in\ca{S}(\ca{H}^S)$ and any $0\leq\lambda\leq1$, it holds that
\alg{
D(\pi\|\lambda\rho+(1-\lambda)\sigma)
\leq\lambda D(\pi\|\rho)+(1-\lambda)D(\pi\|\sigma).
}
\item \label{DMt} Monotonicity: For any unital map $\ca{T}$, it holds that
\alg{
D(\pi\|\rho)\geq D(\pi\|\ca{T}(\rho)).
}
%\item \label{DUI} Unitary Invariance: For any unitary operator $U$, it holds that
%\alg{
%\quad\quad
%D(\pi\|\rho)= D(\pi\|U\rho U^\dagger).
%}
\item \label{DCt} Continuity: $D(\pi\|\rho)$ is a continuous function of $\rho$ in the interior of $\ca{S}(\ca{H}^S)$.
\item \label{DTL} Trace-Log Representation: For any $\rho\in\ca{S}(\ca{H}^S)$, it holds that
\alg{
D(\pi\|\rho)=-\frac{1}{m}{\rm Tr}\ln(m\rho).
\laeq{lavela}
}
\end{enumerate}
\elmm

\bprf
Properties \ref{DLB}$\sim$\ref{DCt} immediately follow from standard properties of the quantum relative entropy.
Property \ref{DTL} follows by
$
-{\rm Tr}\ln(m\rho)
=
-{\rm Tr}\left[\ln(mI)+\ln{\rho}\right]
=
{\rm Tr}\left[\ln{\pi}-\ln{\rho}\right]
=
m{\rm Tr}\left[\pi\ln{\pi}-\pi\ln{\rho}\right]
=
mD(\pi\|\rho)
$. 
\QED
\eprf

\section{Sanov's theorem for continuous variables}
\lsec{Sanovcont}

Let $(\ca{X},\ca{F})$ be a measurable space and let $\Lambda$ be the set of all probability measures on $(\ca{X},\ca{F})$.
A probability measure $P\in\Lambda$ is said to be {\it absolutely continuous with respect to $Q\in\Lambda$} if $P(\ca{B})=0$ for any $\ca{B}\in\ca{F}$ satisfying $Q(\ca{B})=0$, which is represented as $P\ll Q$.
For $P,\:Q\in\Lambda$, the relative entropy (Kullback-Leibler divergence) $D(P\|Q)$ is defined by
\alg{
D(P\|Q)
:=
\begin{cases}
\left.\int_{\ca{X}}\ln{\left(\frac{dP}{dQ}\right)}dP\right.&(P\ll Q)
\\
+\infty&otherwise.
\end{cases}
} 
%Two probability measures $P,Q\in\Lambda$ is said to be mutually absolutely continuous if it holds that $\{\ca{B}\in\ca{F}|P(\ca{B})=0\}=\{\ca{B}\in\ca{F}|Q(\ca{B})=0\}$.
For any subset $\Sigma\subseteq\Lambda$ and any $Q\in\Lambda$, the relative entropy $D(\Sigma\|Q)$ is defined as
\alg{
D(\Sigma\|Q)
:=
\begin{cases}
\inf_{P\in\Sigma}D(P\|Q)&(\Sigma\neq\emptyset)
\\
+\infty&(\Sigma=\emptyset).
\end{cases}
}
The following theorem provides us with a way to calculate $D(\Sigma\|Q)$ when $\Sigma$ is defined in terms of the expectation values of measurable functions:

\bthm{csiszarorig}{\bf[Corollary of Theorem 2 in \cite{csiszar1984sanov}; see also \cite{jupp1983note}] }
Fix $w\in\mbb{N}$ and let $\bm{g}:\ca{X}\rightarrow\mbb{R}^w$ be a measurable function on $(\ca{X},\ca{F})$.
Let $\Sigma\subset\Lambda$ be a convex subset of probability measures on $(\ca{X},\ca{F})$ such that
\alg{
\Sigma
=
\left\{P\in\Lambda\left|\int \bm{g}dP=\bm{0}\right.\right\}.
}
Let $Q\in\Lambda$ be a probability measure such that
$
D(\Sigma\|Q)<\infty
$.
Suppose that there exists a $P\in\Sigma$ that is mutually absolutely continuous with $Q$.
It holds that
\alg{
D(\Sigma\|Q)
=
\max_{\bm{\theta}\in\mbb{R}^w}\left[-\ln{\left\{\int{\exp{\left(\bm{\theta}\cdot\bm{g}\right)}dQ}\right\}}\right].
}
\ethm

\noindent
Let $E$ be a real Hausdorff topological space.
The notion of $\tau$-continuity of a function of probability measures to $E$ was introduced in \cite{groeneboom1979large}(see Section 2 thereof):

\bdfn{taucontinuity}
A function $T:\Lambda\rightarrow E$ is said to be $\tau$-continuous at $P\in\Lambda$ if, for any neighborhood $V\subset E$ of $T(P)$, there exist $\delta>0$ and a partition $\mfk{X}=( \chi_1,\cdots, \chi_K)$ of $\ca{X}$ into sets $ \chi_\kappa\in\ca{F}$, such that $T(Q)\in V$ for any $Q\in\Lambda$ satisfying $\max_{1\leq \kappa\leq K}|Q(\chi_\kappa)-P(\chi_\kappa)|<\delta$.
\edfn

\noindent
Let $X$ be a random variable taking values in $\ca{X}$ according to a probability measure $Q\in\Lambda$.
The empirical distribution $\hat{P}_n^{X^n}$ of an $n$-tuple $X^n=(X_1,\cdots,X_n)$ of $\ca{X}$-valued random variables is the random probability measure defined by
\alg{
\hat{P}_n^{X^n}(\ca{B})
=
\frac{1}{n}|\{i:X_i\in \ca{B}\}|,
\quad
\ca{B}\in\ca{F}.
\laeq{empdist}
}
The following theorem is one of the extensions of Sanov's theorem \cite{sanov1957probability} to general probability spaces:

\bthm{orig}{\bf[Theorem 4.1 in \cite{groeneboom1979large}]}
Let $Q\in\Lambda$ be a probability measure and let $\{\ca{B}_m\in\ca{F}\}_{m=1}^\infty$ be a sequence of Borel sets such that $\ca{B}_m\subset \ca{B}_{m+1}\:(m=1,2,\cdots)$ and $\lim_{m\rightarrow\infty}Q(\ca{B}_m)=1$.
Define $\Lambda_m:=\{P\in\Lambda|P(\ca{B}_m)=1\}$ for $m\in\mbb{N}$ and $\Lambda^*:=\cup_{m=1}^\infty\Lambda_m$.
Let $E$ be a real Hausdorff topological space and let $T:\Lambda^*\rightarrow E$ be a function.
Suppose that the restriction $T|_{\Lambda_m}$ is linear and $\tau$-continuous at each $P\in\Lambda_m$ satisfying $D(P\|Q)<+\infty$ for each $m\in\mbb{N}$.
Let $A$ be a convex subset of $E$ such that $D(T^{-1}({\rm int}A)\|Q)<+\infty$.
It holds that
\alg{
-\lim_{n\rightarrow\infty}\frac{1}{n}\ln{{\rm Pr}\{T(\hat{P}_n^{X^n})\in A\}}
=
D(T^{-1}(A)\|Q).
}
\ethm

We apply \rThm{csiszarorig} and \rthm{orig} to the case where $\ca{X}$ and $E$ is finite-dimensional real vector spaces equipped with the standard topology.
Fix $v,w\in\mbb{N}$.
Let $\Lambda(\mbb{R}^v)$ be the set of all probability measures on $\mbb{R}^v$.
For $P,\:Q\in\Lambda(\mbb{R}^v)$, the relative entropy $D(P\|Q)$ is defined as
\alg{
D(P\|Q)
=
\int_{z\in\mbb{R}^v}P(z)\ln{\frac{P(z)}{Q(z)}}dz.
} 
%which can also be deduced from \req{DPQBorel} (see e.g.~\cite{pinsker1964information}).
The following theorem is immediately obtained as a corollary of \rThm{csiszarorig}:

\bthm{csiszar}
Let $\tilde{\bm{g}}:\mbb{R}^v\rightarrow\mbb{R}^w$ be a continuous function.
For $\bm{t}\in\mbb{R}^w$,
define a subset $\Sigma_{\bm{t}}\subset\Lambda(\mbb{R}^v)$ by
\alg{
\Sigma_{\bm{t}}
=
\left\{P\in\Lambda(\mbb{R}^v)\left|\int_{z\in\mbb{R}^v} P(z)\tilde{\bm{g}}(z)dz=\bm{t}\right.\right\}.
}
Let $Q\in\Lambda$ be a probability measure such that $D(\Sigma_{\bm{t}}\|Q)<+\infty$.
Suppose that there exists a probability measure $P\in\Sigma_{\bm{t}}$ that is mutually absolutely continuous with $Q$. 
Then, it holds that
\alg{
D(\Sigma_{\bm{t}}\|Q)
=
\max_{\bm{\theta}\in\mbb{R}^w}\left[-\ln{\left\{\int_{z\in\mbb{R}^v}{  Q(z)e^{\bm{\theta}\cdot(\tilde{\bm{g}}(z)-\bm{t})}dz}\right\}}\right] . 
}
\ethm

\noindent
Let $Z$ be a random variable taking values in $\mbb{R}^v$ according to a probability measure $Q\in\Lambda$.
In the same way as \req{empdist}, the empirical distribution $\hat{P}_n^{Z^n}$ of an $n$-tuple $Z^n=(Z_1,\cdots,Z_n)$ of $\mbb{R}^v$-valued random variables is the random probability measure defined by
\alg{
\hat{P}_n^{Z^n}(\ca{B})
=
\frac{1}{n}|\{i:Z_i\in \ca{B}\}|,
\quad
\ca{B}\subset\mbb{R}^v.
}
The following theorem is an extension of Sanov's theorem to continuous random variables:

\bthm{sanov}
Let $\bm{f}:\mbb{R}^v\rightarrow\mbb{R}^w$ be a continuous function.
Define a function $\bm{F}:\Lambda(\mbb{R}^v)\rightarrow\mbb{R}^w$ by
\alg{
\bm{F}(P)=\int_{z\in\mbb{R}^v}P(z)\bm{f}(z)dz
}
and $\mfk{D}:\mbb{R}^w\rightarrow\mbb{R}$ by
\alg{
\mfk{D}(\bm{t})
=
D(\bm{F}^{-1}(\{\bm{t}\})\|Q).
\laeq{dfnmfkD}
}
Let $\Xi$ be a subset of $\mbb{R}^w$ such that $\mfk{D}$ is continuous in ${\rm int}\Xi$ and $\mfk{D}(\bm{t'})=+\infty$ for any $\bm{t'}\notin{\rm int}\Xi$.
Let $\Theta$ be a subset of $\Xi$ such that %${\rm cl}\Theta$ is regular and
$D(\bm{F}^{-1}({\rm int}\Theta)\|Q)<+\infty
$.
It holds that
\alg{
-\limsup_{n\rightarrow\infty}\frac{1}{n}\ln{{\rm Pr}\{\hat{P}_n^{Z^n}\in \bm{F}^{-1}(\Theta)\}}
&\geq
\inf_{\bm{t}\in {\rm cl}\Theta}\mfk{D}(\bm{t}),
\laeq{smetana1}
\\
-\liminf_{n\rightarrow\infty}\frac{1}{n}\ln{{\rm Pr}\{\hat{P}_n^{Z^n}\in \bm{F}^{-1}(\Theta)\}}
&\leq
\inf_{\bm{t}\in  {\rm int}\Theta}\mfk{D}(\bm{t}).
\laeq{smetana2}
}
\ethm

\bprf
Follows from \rThm{orig}. See \rApp{prfSanovCont} for a detailed proof.\QED
\eprf

\section{Proof of The Main Result}
\lsec{LDP}

This section provides a proof of \rThm{LDP}.
First, we introduce a few definitions and notations that will be used in the proof.
Second, we prove \rThm{LDP} based on two lemmas and \rThm{sanov} in \rSec{Sanovcont}.
A proof of one of the lemmas regarding the calculation of the exponent will be provided at the end of this section and is based on \rThm{csiszar} in \rSec{Sanovcont}.

\subsection{Definitions and Notations}
\lsec{OPBasis}

Let $\{A_r\}_{r=0}^{m^2-1}$ be a set of Hermitian operators on $\ca{H}^S$ that forms an orthonormal basis of the operator Hilbert space $\ca{L}(\ca{H}^S)$ with respect to the Hilbert-Schmidt inner product, that is, $\frac{1}{m}{\rm Tr}[A_rA_{r'}]=\delta_{rr'}$.
For convenience, we assume that $A_0=I$.
We introduce a notation
\alg{
{\bm A}\equiv(A_0,\cdots,A_{m^2-1}).
}
Define
\alg{
\mbb{R}^{m^2}_{\ca{P}(\ca{H}^S)}:=\{\bm{t}\in\mbb{R}^{m^2}|\bm{t} \cdot \bm{A}\geq0\},
}
where $\bm{t}\cdot\bm{A}\equiv\sum_{r=0}^{m^2-1}t_rA_r$ for $\bm{t}\equiv(t_0,\cdots,t_{m^2-1})\in\mbb{R}^{m^2}$.
The elements of the set $\ca{P}(\ca{H}^S)$ have a one-to-one correspondence with those of $\mbb{R}^{m^2}_{\ca{P}(\ca{H}^S)}$.
Indeed, every element $\varrho$ of $\ca{P}(\ca{H}^S)$ is mapped to an element of $\mbb{R}^{m^2}_{\ca{P}(\ca{H}^S)}$ by $\bm{t}(\varrho)=(1/m){\rm Tr}[\bm{A}\varrho]$, the inverse map of which is given by $\varrho(\bm{t})=\bm{t}\cdot\bm{A}$.
For any subset $\Omega\subseteq\ca{S}(\ca{H}^S)$, we define
\alg{
\mbb{R}^{m^2}_{\Omega}
&:=
\left\{\bm{t}\in\mbb{R}^{m^2}_{\ca{P}(\ca{H}^S)}\backslash\{\bm{0}\}\left|\frac{\bm{t}\cdot\bm{A}}{{\rm Tr}[\bm{t}\cdot\bm{A}]}\in\Omega\right.\right\}.
}
It is straightforward to verify that $\mbb{R}^{m^2}_{\Omega}=\lambda\mbb{R}^{m^2}_{\Omega}$ for any $\lambda>0$, thus $\mbb{R}^{m^2}_{\Omega}$ is a linear cone.
In addition, we have
\alg{
{\rm cl}\mbb{R}^{m^2}_{\Omega}\backslash\{\bm{0}\}=\mbb{R}^{m^2}_{{\rm cl}\Omega},
\quad{\rm int}\mbb{R}^{m^2}_{\Omega}=\mbb{R}^{m^2}_{{\rm int}\Omega}.
\laeq{rararamen}
}

Let $\alpha_r$ and $\beta_r$ be the real part and the imaginary part of $A_r$, i.e.,
\alg{
\alpha_r
:=
\frac{A_r+A_r^*}{2},
\quad
\beta_r
:=
\frac{A_r-A_r^*}{2i}
 \laeq{acroball1}
}
with the superscript $*$ denoting complex conjugate.
Define
$
\alpha_{r,jk}:=\mfk{R}\mfk{e}\bra{j}A_r\ket{k}
$
and
$
\beta_{r,jk}:=\mfk{I}\mfk{m}\bra{j}A_r\ket{k}
$.
Let $ \ca{A}_r\;(r=0,\cdots,m^2-1)$ be a $2m$-dimensional real symmetric matrix defined by
\alg{
 \ca{A}_r
:=
\begin{pmatrix}
\alpha_r & \beta_r \\
-\beta_r & \alpha_r
\end{pmatrix}
=
\left(
\begin{array}{c|c} 
  \left[\alpha_{r,jk}\right]_{jk} &   \left[\beta_{r,jk}\right]_{jk}  \\ \hline
  -\left[\beta_{r,jk}\right]_{jk} & \left[\alpha_{r,jk}\right]_{jk} 
\end{array}
\right).
\laeq{custom1}
}
We introduce a notation
\alg{
\bm{\ca{A}}\equiv(\ca{A}_0,\cdots,\ca{A}_{m^2-1}).
}
A simple calculation yields
\alg{
\ca{A}_r
&
=
I\otimes \alpha_r+i\sigma_y\otimes  \beta_r
\\
&=
\frac{I+\sigma_y}{2}\otm (  \alpha_r+i \beta_r)
+
\frac{I-\sigma_y}{2}\otm (  \alpha_r-i \beta_r)
\\
&=
\proj{\varphi_+}\otm A_r
+
\proj{\varphi_-}\otm A_r^*,
\laeq{hontaik}
}
where $\varphi_\pm$ are the eigenvectors of $\sigma_y$.
It follows that
\alg{
\bm{\ca{A}}
=
\proj{\varphi_+}\otm \bm{A}
+
\proj{\varphi_-}\otm \bm{A}^*.
\laeq{hontai}
}

Let $\Lambda(\mbb{R}^{2m})$ be the set of all probability measures on $\mbb{R}^{2m}$.
Define functionals $\Phi_r:\Lambda(\mbb{R}^{2m})\rightarrow\mbb{R}\:(r=0,\cdots,m^2-1)$ and $\bm\Phi:\Lambda(\mbb{R}^{2m})\rightarrow\mbb{R}^{m^2}$ by
\alg{
\Phi_r(P)
:=
\int_{\mbb{R}^{2m}} P(z)z^T\ca{A}_r zdz
}
and
\alg{
\bm\Phi(P)
:=
(\Phi_0(P),\cdots,\Phi_{m^2-1}(P)).
\laeq{dansei}
}
It is straightforward to verify that $\bm{\Phi}$ satisfies the condition of $\bm{F}$ in \rThm{sanov}.

The one-dimensional standard normal distribution is defined by the probability density
\alg{
G_1(z):=\frac{1}{\sqrt{2\pi}}e^{-\frac{1}{2}z^2}\quad(z\in\mbb{R}).
\laeq{dfnSND}
}
Likewise, the standard normal distribution on the $D$-dimensional real vector space is defined by the probability density
\alg{
G_D(z):=\frac{1}{\sqrt{(2\pi)^D}}e^{-\frac{1}{2}z^Tz}\quad(z\in\mbb{R}^D),
}
which is equivalent to
$
G_D(z)=\prod_{d=1}^DG_1(z_d)
$
for $z\equiv(z_1,\cdots,z_D)$.
%In general, a multivariate normal distribution with mean zero is defined by a positive semidefinite real matrix $\ca{A}$ as
%\alg{
%G_M^A(\Phi)=\frac{1}{\sqrt{(2\pi)^M|\ca{A}|}}e^{-\frac{1}{2}\Phi^T\ca{A}\Phi}\quad(\Phi\in\mbb{R}^M).
%}

\subsection{Proof of \rThm{LDP}}
\lsec{parsii}

We prove \rThm{LDP} based on the two lemmas presented below.
The first one is that the unitary invariant measure on the set of normalized state vectors are generated by the multivariate standard normal distribution:

\blmm{unitaryinvariantmeasure}
Let $\{X_d\}_{d=1}^D$ and $\{Y_d\}_{d=1}^D$ be a collection of independent and identically distributed random variables that obey the standard normal distribution.
Let $\{|e_d\rangle\}_{d=1}^D$ be a fixed orthonormal basis of $\mbb{C}^D$, and let $\ca{S}(\mbb{C}^D)\subset\mbb{C}^D$ be the set of unit vectors in $\mbb{C}^D$. 
Define a vector-valued random variable $Z$, taking values in $\mbb{C}^D$, as
$
Z=\sum_{d=1}^D(X_d+iY_d)\ket{e_d}.
$
Let $\mu:{\rm Borel}(\ca{S}(\mbb{C}^D))\rightarrow[0,1]$ be the Borel probability measure on $\ca{S}(\mbb{C}^D)$ defined by
$
\mu(\ca{B})={\rm Pr}\{\lambda Z\in\ca{B}\text{ for some }\lambda>0\}
$.
The measure $\mu$ is unitarily invariant, i.e., $\mu(U\ca{B})=\mu(\ca{B})$ for any $\ca{B}\in{\rm Borel}(\ca{S}(\mbb{C}^D))$ and any unitary matrix $U$ on $\mbb{C}^D$.
\elmm

\bprf
See e.g.~Section 7.2 in \cite{watrous2018theory}.
\QED
\eprf

\noindent
The second one is that the infimum of the relative entropy between the set $\Phi^{-1}(\{\bm{t}\})$ and the standard normal distribution is represented by the unnormalized density matrix corresponding to $\bm{t}$:

\blmm{bokuj}
For any $\bm{t}\in\mbb{R}^{m^2}$, define
\alg{
\mfk{D}(\bm{t})
:=
D(\bm{\Phi}^{-1}(\{\bm{t}\})\|G_{2m}).
\laeq{oruoru11}
}
It holds that
\alg{
\mfk{D}(\bm{t})=
\begin{cases}
{\rm Tr}
\left[
\varsigma(\bm{t}) - I 
- \ln{(\varsigma(\bm{t}))}
\right] &(\bm{t}\in{\rm int}\mbb{R}^{m^2}_{\ca{P}(\ca{H}^S)})
\\
+\infty &(\bm{t}\notin{\rm int}\mbb{R}^{m^2}_{\ca{P}(\ca{H}^S)})
\end{cases},
\laeq{grievous}
}
where $\varsigma(\bm{t}):=\bm{t}\cdot\bm{A}/2m$.
Thus, $\mfk{D}(\bm{t})$ is continuous in $\mbb{R}^{m^2}_{\ca{P}(\ca{H}^S)}$.
\elmm

\bprf
See \rSec{prfbokuj}.
\QED
\eprf

\noindent
{\bf Proof of \rThm{LDP}:}
Let $\{\ket{k}\}_{k=1}^m$ and $\{\ket{l}\}_{l=1}^n$ be a fixed orthonormal basis of $\ca{H}^S$ and $\ca{H}^E$, respectively.
Let the orthonormal basis of the whole space $\ca{H}=\ca{H}^S\otm\ca{H}^E$ be $\{\ket{e_{kl}}\}_{1\leq k\leq m, 1\leq l\leq n}$, where $\ket{e_{kl}}=\ket{k}\ket{l}$.
Let $X_{kl}$ and $Y_{kl}$ ($1\leq k\leq m, 1\leq l\leq n$) be real random variables that are independently and identically distributed with respect to the standard normal distribution (SND).
Let $\ket{\tilde{\Psi}}$ be a random unnormalized vector defined by
\alg{
\ket{\tilde{\Psi}}
=
\sum_{\substack{1\leq k\leq m\\1\leq l\leq n}}(X_{kl}+iY_{kl})\ket{e_{kl}}.
\laeq{tadori}
}
With a slight abuse of notation, we denote $\tilde{\Psi}\sim{\rm SND}(n)$.
Due to \rLmm{unitaryinvariantmeasure}, the random state $\ket{\Psi}$ distributed with respect to the unitarily invariant measure on $\ca{H}$ is equivalent to the random state $\ket{\tilde{\Psi}}/\|\ket{\tilde{\Psi}}\|$.
With $\varrho_{\tilde{\Psi}}$ denoting ${\rm Tr}_E[|\tilde{\Psi}\rangle\!\langle\tilde{\Psi}|]$,
we have the following equivalence relations:
\alg{
{\rm Tr}_E\left[\frac{\proj{\tilde{\Psi}}}{\|\ket{\tilde{\Psi}}\|^2}\right]\in\Omega
\;\Leftrightarrow\;
\frac{\varrho_{\tilde{\Psi}}}{{\rm Tr}[\varrho_{\tilde{\Psi}}]}\in\Omega
\;\Leftrightarrow\;
{\rm Tr}[\bm{A}\varrho_{\tilde{\Psi}}]\in\mbb{R}^{m^2}_{\Omega}.
}
Thus, noting that ${\rm Tr}[{\bm A}\varrho_{\tilde{\Psi}}]=\langle\tilde{\Psi}|({\bm A}^S \otm  I^E)|\tilde{\Psi}\rangle$, we have
\alg{
{\rm Pr}\left\{\rho_\Psi\in\Omega|\Psi \sim \mu_n\right\}
=
{\rm Pr} \left\{\left.  \bra{\tilde{\Psi}}({\bm A}^S \otm  I^E)\ket{\tilde{\Psi}} \in \mbb{R}^{m^2}_{\Omega}\right|\tilde{\Psi}  \sim   \text{SND}(n) \right\}  . 
\laeq{NGC}
}

To evaluate the probability \req{NGC}, we introduce a $2m$-dimensional real vector random variable
\alg{
Z=(X_1,\cdots ,X_k,\cdots ,X_m; Y_1,\cdots ,Y_{k'},\cdots ,Y_m)^T.
}
Here, $X_k$ and $Y_{k'}$ are distributed independently with respect to the standard normal distribution.
I.e., $Z$ obeys a probability distribution
\alg{
G_{2m}(Z)
=
\prod_{k=1}^mG_1(X_k)G_1(Y_k).
}
Let $Z^n\equiv(Z_1,\cdots,Z_l,\cdots,Z_n)$ be an independently and identically distributed sequence of the random variable $Z$, where
\alg{
  Z_l
\equiv
(X_{1l},\cdots ,X_{kl},\cdots ,X_{ml}; Y_{1l},\cdots ,Y_{k'l},\cdots ,Y_{ml})^T .  
}
Noting that $\beta_{r,jk}=-\beta_{r,kj}$, we have
\alg{
&\bra{\tilde{\Psi}}( A_r^S\otm I^E)\ket{\tilde{\Psi}}
\nn\\
&
=
\sum_{1\leq l\leq n}\left(\sum_{\substack{1\leq j\leq m\\1\leq k\leq m}}\alpha_{r,jk}(X_{jl}X_{kl}+Y_{jl}Y_{kl})\right)
+\sum_{1\leq l\leq n}\left(
\sum_{\substack{1\leq j\leq m\\1\leq k\leq m}}\beta_{r,jk}(X_{jl}Y_{kl}-Y_{jl}X_{kl})
\right)
\\
&=\sum_{1\leq l\leq n}Z_l^T \ca{A}_r Z_l.
\laeq{aXYklmulti}
}
Thus, the probability \req{NGC} is equal to
\alg{
{\rm Pr}\left\{\left. 
\frac{1}{n}
\sum_{1\leq l\leq n}Z_l^T\bm{\ca{A}} Z_l
\in\mbb{R}^{m^2}_{\Omega}\right|Z_l \sim  \text{SND} \right\}
=
{\rm Pr}\left\{ \hat{P}_n^{Z^n}
\in\bm{\Phi}^{-1}(\mbb{R}^{m^2}_{\Omega})\right\},
}
where $\hat{P}_n^{Z^n}$ denotes the empirical distribution of the sequence.

We apply \rThm{sanov}, which is possible due to \rLmm{bokuj}.
It follows that
\alg{
 -\limsup_{n\rightarrow\infty}
\frac{1}{n}\ln{{\rm Pr}\left\{ \hat{P}_n^{Z^n}
\in\bm{\Phi}^{-1}(\mbb{R}^{m^2}_{\Omega})\right\}}
&\geq
\inf_{{\bm t}\in{\rm cl}\mbb{R}^{m^2}_{\Omega}}\mfk{D}({\bm t}),
\\
 -\liminf_{n\rightarrow\infty}
\frac{1}{n}\ln{{\rm Pr}\left\{ \hat{P}_n^{Z^n}
\in\bm{\Phi}^{-1}(\mbb{R}^{m^2}_{\Omega})\right\}}
&\leq
\inf_{{\bm t}\in{\rm int}\mbb{R}^{m^2}_{\Omega}}\mfk{D}({\bm t}).
}
For both $\ast={\rm cl},{\rm int}$, the infimum is calculated as
\alg{
\inf_{{\bm t}\in\ast\mbb{R}^{m^2}_{\Omega}}\mfk{D}({\bm t})
&
=
\inf_{\bm{t}\in\ast\mbb{R}^{m^2}_{\Omega}}
{\rm Tr}
\left[
\frac{\bm{t}\cdot\bm{A}}{2m} - I 
- \ln{\left(\frac{\bm{t}\cdot\bm{A}}{2m}\right)}
\right]
\\
&
=
 \inf_{\bm{t}\in\ast\mbb{R}^{m^2}_{\Omega}}
 \left[m \left(\frac{t_0}{2m} - 1 - \ln{\frac{t_0}{2m}}\right) - {\rm Tr}\ln{\left(\frac{\bm{t}\cdot\bm{A}}{t_0}\right)}\right] 
\laeq{kasuki}\\
&
=
\inf_{\bm{t}\in\ast\mbb{R}^{m^2}_{\Omega}}
-{\rm Tr}\ln{\left(\frac{\bm{t}\cdot\bm{A}}{t_0}\right)}
\laeq{habukish}\\
&
=
\inf_{\rho\in\ast\Omega}
-{\rm Tr}\ln{(m\rho)}
\laeq{sarus}\\
&
=
m\inf_{\rho\in\ast\Omega}
D(\pi\|\rho).
\laeq{suteki}
}
Here, \req{habukish} follows from $\inf_{t_0>0}\{t_0/2m-1-\ln{(t_0/2m)}\}=0$ and the fact that $\ast\mbb{R}^{m^2}_{\Omega}$ is a linear cone;
\req{sarus} from \req{rararamen} the fact that the condition $\bm{t}\in\ast\mbb{R}^{m^2}_{\Omega}$ is equivalent to $\bm{t}\cdot\bm{A}/mt_0\in\ast\Omega$ because ${\rm Tr}[\bm{t}\cdot\bm{A}]=mt_0$;
and the last line due to \req{lavela}.
%By assumption, we have that $\inf_{\rho\in{\rm int}\Omega}D(\pi\|\rho)=\inf_{\rho\in{\rm cl}\Omega}D(\pi\|\rho)$.
All in all, we obtain \req{gojupe} and \req{gojure}.

When $\Omega$ satisfies ${\rm cl}({\rm int}\Omega)={\rm cl}\Omega$, it holds that
\alg{
\inf_{\rho\in{\rm cl}\Omega}D(\pi\|\rho)
=
\inf_{\rho\in{\rm int}\Omega}D(\pi\|\rho)
}
due to the continuity of $D(\pi\|\rho)$.
This implies \req{reibou} and complete the proof of \rThm{LDP}.
\QED

\subsection{Proof of \rLmm{bokuj}}
\lsec{prfbokuj}

Let $z\equiv(z_1,\cdots,z_{2m})^T$ denote the concrete values that $Z$ takes.
By definition, the condition $P\in\bm{\Phi}^{-1}(\{\bm{t}\})$ is equivalent to
\alg{
\int_{z\in\mbb{R}^{2m}} P(z)z^T\bm{\ca{A}} zdz
=\bm{t}.
\laeq{mala}
}

We first consider the case where $\bm{t}\in{\rm int}\mbb{R}^{m^2}_{\ca{P}(\ca{H}^S)}$, which is equivalent to $\bm{t}\cdot\bm{A}>0$.
Define $\bm{\theta}^*(\bm{t})\in\mbb{R}^{m^2}$ by
\alg{
\bm{\theta}^*(\bm{t})
:=\frac{{\rm Tr}{\bm A}}{2m}-{\rm Tr}[{\bm A}(\bm{t}\cdot{\bm A})^{-1}].
\laeq{kansou}
}
Due to the complete orthonormality of $\{A_r\}_{r=0}^{m^2-1}$, the expression \req{kansou} is equivalent to 
\alg{
I-2\bm{\theta}^* (\bm{t}) \cdot \bm{ A}=\left(\frac{\bm{t}\cdot\bm{A}}{2m}\right)^{-1},
\laeq{dodor}
}
which in turn is equivalent to
%&
%\bm{t}=
%2{\rm Tr}\left[ \bm{ A}(I-2 \bm{\theta}^*(\bm{t}) \cdot  \bm{A})^{-1}\right],
%\laeq{otuotu1}
%\\
\alg{
\frac{\partial}{\partial \bm{\theta}}
\left.\left[\bm{\theta}\cdot\bm{t}+{\rm Tr}\ln{
\left(I-2\bm{\theta}\cdot\bm{A}\right)}\right]\right|_{\bm{\theta}=\bm{\theta}^*(\bm{t})}
=
\bm{0}.
\laeq{dakuda}
}
Define a function $\mfk{L}:\mbb{R}^{m^2}\times\mbb{R}^{m^2}\rightarrow\mbb{R}$ by
\alg{
\mfk{L}(\bm{\theta},\bm{t})
=
\int_{z\in\mbb{R}^{2m}}{G_{2m}(z)\exp{\left(\bm{\theta} \cdot (z^T\bm{\ca{A}} z-\bm{t})\right)}dz}.
}
With $\ca{I}$ denoting the $2m\times2m$ identity matrix,
we have
\alg{
G_{2m}(z)\exp{\left(\bm{\theta} \cdot (z^T\bm{\ca{A}} z-\bm{t})\right)}
=
\frac{1}{(2\pi)^m}\exp{ \left(-\bm{\theta} \cdot \bm{t}-\frac{1}{2}z^T(\ca{I}-2\bm{\theta} \cdot \bm{\ca{A}})z \right)}.
}
Thus, by taking the Gaussian integral, we obtain
\alg{
\mfk{L}(\bm{\theta},\bm{t})
=
\begin{cases}
\exp{(-\bm{\theta} \cdot \bm{t})}({\rm det}(\ca{I}-2\bm{\theta} \cdot \bm{\ca{A}}))^{-1/2} & (\ca{I}-2\bm{\theta} \cdot \bm{\ca{A}}>0)
\\
+\infty & \text{otherwise}.
\end{cases}
\laeq{ababu}
}
From \req{hontai}, it holds that
\alg{
\ca{I}-2\bm{\theta}\cdot\bm{\ca{A}}
=
\proj{\varphi_+}\otm (I-2\bm{\theta}\cdot\bm{A})
+
\proj{\varphi_-}\otm (I-2\bm{\theta}\cdot\bm{A}^*).
}
Hence, \req{ababu} yields
\alg{
-\ln\mfk{L}(\bm{\theta},\bm{t})
=
\begin{cases}
\bm{\theta} \cdot \bm{t}+{\rm Tr}\ln(I-2\bm{\theta} \cdot \bm{A}) & (I-2\bm{\theta} \cdot \bm{A}>0)
\\
-\infty & \text{otherwise}.
\end{cases}
\laeq{relando}
}
Due to \req{dakuda} and the concavity of the log function, we obtain\alg{
\bm{\theta}^*(\bm{t})
=
\argmax_{\bm{\theta}\in\mbb{R}^{m^2}}\left[-\ln{\mfk{L}(\bm{\theta},\bm{t})}\right].
\laeq{sorando}
}
From \req{kansou}, we have
\alg{
\bm{\theta}^*(\bm{t}) \cdot \bm{t}
=
{\rm Tr}\left[\frac{\bm{t}\cdot\bm{A}}{2m}-I\right].
\laeq{patrol}
}
Substituting this and \req{dodor} to \req{relando}, and noting that we have $I-2\bm{\theta}^* (\bm{t}) \cdot \bm{ A}>0$ due to 
\req{dodor}, we arrive at
\alg{
\max_{\bm{\theta}\in\mbb{R}^{m^2}}\left[-\ln{\mfk{L}(\bm{\theta},\bm{t})}\right]
=
-\ln{\mfk{L}(\bm{\theta}^*(\bm{t}),\bm{t})}
=
{\rm Tr}\left[\frac{\bm{t}\cdot\bm{A}}{2m}-I-\ln{\left(\frac{\bm{t}\cdot\bm{A}}{2m}\right)}\right].
\laeq{oruorai}
}
Define a probability measure $P^*$ on $\mbb{R}^{2m}$ by
\alg{
P^*(z)
=
\frac{G_{2m}(z)\exp{\left[\bm{\theta}^*(\bm{t}) \cdot (z^T\bm{\ca{A}} z-\bm{t})\right]}}{\mfk{L}(\bm{\theta}^*(\bm{t}),\bm{t})}.
}
It is obvious that $P^*(z)>0$ for any $z\in\mbb{R}^{2m}$.
In addition, we have
\alg{
\int_{z\in\mbb{R}^{2m}} P^*(z)(z^T\bm{\ca{A}} z-\bm{t})dz
=\left.\frac{\partial}{\partial \bm{\theta}}\ln\mfk{L}(\bm{\theta},\bm{t})\right|_{\bm{\theta}=\bm{\theta}^*(\bm{t})}=\bm{0},
\laeq{raiku}
}
which implies $P^*\in\bm{\Phi}^{-1}(\{\bm{t}\})$.
Furthermore, a simple calculation using \req{raiku} yields
\alg{
D(P^*\|G_{2m})
=
-\ln{\mfk{L}(\bm{\theta}^*(\bm{t}),\bm{t})}
<+\infty.
}
Thus, it follows from \rThm{csiszar} and \req{oruorai} that
\alg{
D(\bm{\Phi}^{-1}(\{\bm{t}\})\|G_{2m})
=
{\rm Tr}\left[\varsigma(\bm{t})-I-\ln(\varsigma(\bm{t}))\right].
\laeq{sasis}
}

It remains to consider the case where $\bm{t}\notin{\rm int}\mbb{R}^{m^2}_{\ca{P}(\ca{H}^S)}$.
For each $z\in\mbb{R}^{2m}$, define an unnormalized vector $|z\rangle\in\ca{H}^S$ by
\alg{
|z\rangle
:=
\sum_{l=1}^m(z_l-iz_{m+l})|e_l\rangle.
}
Similarly to \req{hontaik}, we have 
$
z=|\varphi_+\rangle|z\rangle
+
|\varphi_-\rangle|z^*\rangle
$.
Thus, the condition \req{mala} is equivalent to
%\alg{
%2\int_{z\in\mbb{R}^{2m}} P(z)\langle z|\bm{A}|z\rangle dz
%=\bm{t},
%}
%which yields
\alg{
\int_{z\in\mbb{R}^{2m}} P(z)\proj{z} dz
=\frac{\bm{t}\cdot\bm{A}}{m}.
\laeq{sosoo}
}
For $\bm{t}\notin{\rm cl}\mbb{R}^{m^2}_{\ca{P}(\ca{H}^S)}$, at least one of the eigenvalues of $\bm{t}\cdot\bm{A}$ is negative.
Thus, no probability measure $P$ satisfies \req{sosoo}, which implies $\mfk{D}(\bm{t})=+\infty$.
For $\bm{t}\in\partial\mbb{R}^{m^2}_{\ca{P}(\ca{H}^S)}$, at least one of the eigenvalues of $\bm{t}\cdot\bm{A}$ is equal to zero.
Thus, there exists a linear subspace $\mbb{V}_{\bm{t}}\subsetneq\mbb{R}^{2m}$ such that ${\rm supp}P\subset\mbb{V}_{\bm{t}}$ for any $P$ satisfying \req{sosoo}.
%Thus implies that, with an appropriate orthogonal transformation of $z$ into $z'$, any $P$ satisfying \req{sosoo} can be written as $P(z')=\delta(z_l')\tilde{P}(z_r')$, where $z_l'=(z_1',\cdots,z_{{\rm dim}\mbb{V}_{\bm{t}}})$, $z_r'=(z_{{\rm dim}\mbb{V}_{\bm{t}}+1},\cdots,z_{2m})$, $z'=(z_1',\cdots,z_{2m}')$ and $\delta$ denotes Dirac's delta function.
This implies $\mfk{D}(\bm{t})=+\infty$ and completes the proof.
\QED

\section{Applications }
\lsec{examples}

We apply \rThm{LDP} to a few cases for which the exponent can be calculated concretely.

\subsection{Trace Distance}

We consider the set of states such that the trace distance with the maximally mixed state $\pi$ is no smaller than $t$, where $0< t<1$.
We prove that
\alg{
\inf_{\rho:\frac{1}{2}\|\pi-\rho\|_1= t}D(\pi\|\rho)
=
\mathscr{D}_{\rm TD}(t;m)
:=\min_{m'}\mathscr{D}\left(\left.\frac{m'}{m}\right\|\frac{m'}{m}+t\right),
\laeq{yamayama}
}
where the minimum is taken over $m'\in\{1,\cdots,\lfloor m(1-t)\rfloor\}$.
Since $\mathscr{D}_{\rm TD}(t;m)$ is monotonically non-decreasing in $t$, it follows that
\alg{
\inf_{\rho:\frac{1}{2}\|\pi-\rho\|_1\geq t}D(\pi\|\rho)
=
\mathscr{D}_{\rm TD}(t;m).
\laeq{sannnenn}
}
%A graph of this function is depicted in \rFig{}.

First, we prove that 
$
D(\pi\|\rho)
\geq
\mathscr{D}_{\rm TD}(t;m)
$
for any $\rho$ satisfying $\|\pi-\rho\|_1/2= t$.
Let $\Pi_\rho$ be the projection of the subspace of $\ca{H}^S$ spanned by the eigenvectors of $\rho-\pi$ with nonnegative eigenvalues.
Define probability distributions $P\equiv\{p(0),p(1)\}$ and $Q\equiv\{q(0),q(1)\}$ by
\alg{
p(0)={\rm Tr}[\Pi_\rho\rho],
\;
p(1)={\rm Tr}[(I-\Pi_\rho)\rho],
\;q(0)={\rm Tr}[\Pi_\rho\pi],
\;
q(1)={\rm Tr}[(I-\Pi_\rho)\pi].
}
Due to the property of the trace distance (see e.g.~Section 9.1.3 of \cite{wildetext}), we have
\alg{
p(0)-q(0)=q(1)-p(1)=t.
}
Letting $m'\equiv{\rm rank}\Pi_\rho$, it follows that
\alg{
q(0)=\frac{m'}{m},
\;
q(1)=1-\frac{m'}{m},
\;
p(0)=\frac{m'}{m}+t,
\;
p(1)=1-\frac{m'}{m}-t.
}
Thus, due to the monotonicity of the quantum relative entropy, we obtain
$
D(\pi\|\rho)
\geq
D(P\|Q)
\geq
\mathscr{D}_{\rm TD}(t;m)
$.

Second, we prove that  there exists a state $\varrho$ such that $\|\pi-\varrho\|_1/2= t$ and
$
D(\pi\|\varrho)
=
\mathscr{D}_{\rm TD}(t;m)
$.
Define
\alg{
m^*_t:=\argmin_{m'}\mathscr{D}\left(\left.\frac{m'}{m}\right\|\frac{m'}{m}+t\right),
}
where $m'$ runs over $\{1,\cdots,\lfloor m(1-t)\rfloor\}$.
Let $\Pi_{m^*_t}$ be an arbitrary projection of rank $m^*_t$.
Consider a state
\alg{
\varrho_t^*
=
\left(\frac{m^*_t}{m}+t\right)\frac{\Pi_{m^*_t}}{m^*_t}
+
\left(1-\frac{m^*_t}{m}-t\right)\frac{I-\Pi_{m^*_t}}{m-m^*_t}.
}
It is straightforward to verify that $D(\pi\|\varrho_t^*)=\mathscr{D}_{\rm TD}(t;m)$.

We consider the case where $m\gg1$ and $t\ll1$.
Due to the properties of the binary relative entropy (see \rApp{BRE}), it follows from \req{yamayama} and \req{sannnenn} that
\alg{
\inf_{\rho:\frac{1}{2}\|\pi-\rho\|_1= t}D(\pi\|\rho)
\approx
\min_{\alpha}\mathscr{D}\left(\left.\alpha\right\|\alpha+t\right)
\approx
\frac{t^2}{2}.
}

\subsection{Entropy}
\lsec{RFentropy}

\begin{figure}[t]
\begin{center}
\includegraphics[bb={10 25 292 220}, scale=1.3]{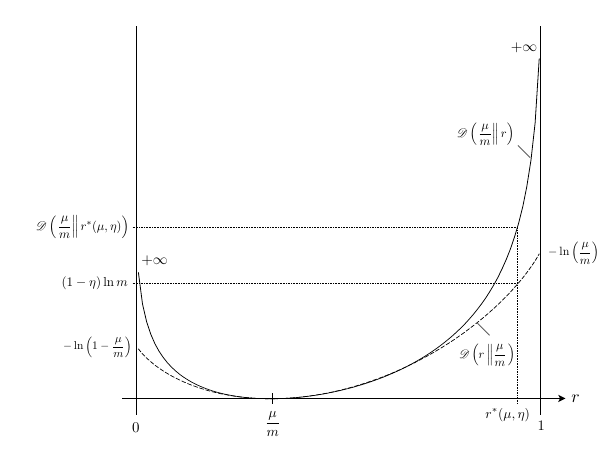}
\end{center}
\caption{
This graph shows the way how $\mathscr{D} \left( \left.\mu/m\right\|r^*(\mu,\eta) \right)$ is obtained for a fixed $m$, $\mu$ and $\eta$. 
The solid line and the dashed line represents the functions $\mathscr{D} \left( \left.\mu/m\right\|r\right)$ and $\mathscr{D} \left( r\left\|\mu/m\right. \right)$, respectively.
}
\label{fig:entropygraph}
\end{figure}

The von Neumann entropy of a state $\rho\in\ca{S}(\ca{H}^S)$ is defined by
\alg{
H(\rho)
:=-{\rm Tr}[\rho\ln{\rho}].
}
We prove that
\alg{
 \inf_{\rho:H(\rho)= \eta\ln{m}}  D(\pi\|\rho)
=\mathscr{D}_{\rm vN}(\eta;m)
:=\min_{\mu}\mathscr{D} \left( \left.\frac{\mu}{m}\right\|r^*(\mu,\eta) \right)  
\laeq{maiza}
}
for any $0<\eta<1$, where the minimum is taken over $\mu\in\{1,\cdots,m-1\}$.
Here, $r^*(\mu,\eta)$ is the solution of an equation 
\alg{
\mathscr{D}\left(r\left\|\frac{\mu}{m}\right.\right)=(1-\eta)\ln{m}.
\laeq{acac}
}
A graphical representation of the function $\mathscr{D} \left( \left.\frac{\mu}{m}\right\|r^*(\mu,\eta) \right)$ is presented in \rFig{entropygraph}.
Due to the convexity of the relative entropy, the equation \req{acac} has at most two solutions.
Since \req{maiza} and \req{acac} are both invariant under the simultaneous exchange of $\mu$ with $m-\mu$ and $r$ with $1-r$, we may assume $r^*(\mu,\eta)$ to be the smaller one of the solutions.
Due to the property of the binary relative entropy (see \rApp{BRE}), $\mathscr{D}_{\rm vN}(\eta;m)$ is monotonically non-increasing in $\eta$.
Thus, it holds that
\alg{
\inf_{\rho:H(\rho)\leq \eta\ln{m}}D(\pi\|\rho)
=\mathscr{D}_{\rm vN}(\eta;m).
}

To prove \req{maiza}, note that both $H(\rho)$ and $D(\pi\|\rho)$ depend only on the spectrum of $\rho$.
Thus, it is sufficient to consider the case where $\pi$ and $\rho$ are classical probability distributions.
That is, we only need to prove that
\alg{
\inf_{P:H(P)= \eta\ln{m}}D(U\|P)
=\mathscr{D}_{\rm vN}(\eta;m),
\laeq{maihagi}
}
where the infimum is taken over all probability distributions $P$ on the set $\{1,\cdots,m\}$ and $U$ is the uniform distribution thereon.
Since the set of probability distributions satisfying $H(P)= \eta\ln{m}$ is closed, the infimum in \req{maihagi} can be replaced by the minimum.
Hence, denoting $P\equiv\{p(1),\cdots,p(m)\}$, we are to minimize the function
\alg{
D(U\|P)
=
-\ln{m}-\frac{1}{m}\sum_{k=1}^m\ln{p(k)},
\laeq{sssa}
}
under the condition that
\alg{
\sum_{k=1}^mp(k)=1,
\quad
-\sum_{k=1}^mp(k)\ln{p(k)}=\eta\ln{m}.
\laeq{fufuse}
}
We apply the Lagrange multiplier method.
Let $P_*\equiv\{p_*(1),\cdots,p_*(m)\}$ be
\alg{
P_*:=\argmin_{P:H(P)= \eta\ln{m}}D(U\|P).
}
With $u$ and $v$ being the Lagrange multipliers, the condition for the extremal value is given by
\alg{
\frac{1}{p_*(k)}
-u(1+\ln{p_*(k)})
-v=0
\;
(k=1,\cdots,m). 
}
Due to the concavity of the logarithm function, every equation of the form
\alg{
x+C_1\ln{x}+C_2=0
\quad
(C_1,C_2\in\mbb{R})
}
has at most two solutions.
Thus, $P_*$ is of the form
\alg{
p(k)
=
\begin{cases}
q_0&(1\leq k\leq\mu)\\
q_1&(\mu+1\leq k\leq m)
\end{cases},
\laeq{guagua}
}
where $\mu\in\{1,\cdots,m\}$ and $q_0$ and $q_1$ satisfy the following conditions that follow from \req{fufuse}:
\alg{
\mu q_0+(m-\mu)q_1&=1,
\laeq{senden}
\\
-\mu q_0\ln{q_0}-(m-\mu)q_1\ln{q_1}&=\eta\ln{m}.
\laeq{senren}
}
Solving \req{senden} with respect to $q_1$ and substituting it to \req{senren}, we obtain
\alg{
-\mu q_0\ln{\left(\frac{\mu q_0}{\mu}\right)}-(1-\mu q_0)\ln{\left(\frac{1-\mu q_0}{m-\mu}\right)}=\eta\ln{m}.
}
Letting $r\equiv\mu q_0$, this is equivalent to \req{acac}.
From \req{guagua}, the relative entropy \req{sssa} is calculated as
\alg{
D(U\|P)
=
\frac{\mu}{m} \ln {\left( \frac{\mu/m}{\mu q_0} \right)}
 + 
\left(1 - \frac{\mu}{m}\right) \ln {\left( \frac{1 - \mu/m}{1-\mu q_0} \right)}, 
}
which is equal to $\mathscr{D}(\mu/m\|r)$.
Minimizing over $\mu$, we obtain \req{maiza}.

We consider the case where $\eta=1-\delta$, $\delta\ll1$ and $m\gg1$.
As in \req{maiza} and \req{acac}, we are to minimize the function $\mathscr{D}(\alpha\|r)$
over $0<\alpha<1$ and $0<r<1$, under the condition that
\alg{
\mathscr{D}(r\|\alpha)=\delta\ln{m}.
\laeq{perierie}
}
Due to the properties of the binary relative entropy (see \rApp{BRE}), we have
$
\mathscr{D}(\alpha\|\alpha+\epsilon)\approx \mathscr{D}(\alpha+\epsilon\|\alpha)
$
in the second order of $\epsilon\ll0$.
Substituting $r=\alpha+\epsilon$ to \req{perierie} and $\mathscr{D}(\alpha\|r)$, we arrive at
\alg{
\left.\mathscr{D}_{\rm vN}(\eta;m)\right|_{\eta=\delta\ln{m}}
\approx
\min_{\substack{0<\alpha<1\\0<r<1}}\mathscr{D}(\alpha\|r)
\approx
\delta\ln{m}
}
in the first order of $\delta$.
As an upper bound on the probability of large deviation, this is a quadratic improvement with respect to $\delta$ and an improvement by factor $8\pi^2\ln{m}$ over the exponent obtained from Theorem III.3 in \cite{hayden2006aspects} (see \rApp{compPRPR} for the detail).

\subsection{Expectation Value}
\lsec{expvalue}

Let $W$ be a Hermitian operator on $\ca{H}^S$. 
For $w\in\mbb{R}$, we calculate the exponent for the set of states such that the expectation value of $W$ is equal to $w$.
Without loss of generality, we may assume that ${\rm Tr}[W]=0$ and $0<w<\|W\|_\infty$.
We prove that
\alg{
\inf_{\rho:{\rm Tr}[W\rho]=w}D(\pi\|\rho)
=
\mathscr{D}_{\rm exp}(w;W,m)
:=
\frac{1}{m}{\rm Tr}\ln\left[\left(1-w\nu^*(w)\right)I+\nu^*(w) W\right],
\laeq{houjicha}
}
where $\nu^*(w)$ is the solution of 
\alg{
\frac{1}{m}{\rm Tr}[\{(1-w\nu)I+\nu W\}^{-1}]
=1
\laeq{houjisha}
}
under the condition that $\nu\neq0$ and
\alg{
\frac{1}{m}
\{(1-w\nu)I+\nu W\}^{-1}
\geq0.
\laeq{houjijha}
}

Let $\{A_r\}_{r=1}^{m^2-1}$ be the set of traceless Hermitian operators on $\ca{H}$ that are orthonormal with respect to the Hilbert-Schmidt inner product. Every state $\rho\in\ca{S}(\ca{H})$ is represented by a $(m^2-1)$-dimensional real vector $\vec{t}\equiv(t_1,\cdots,t_{m^2-1})\in\mbb{R}^{m^2-1}$ as
\alg{
\rho
=\rho(\vec{t})
:=
\frac{1}{m}(I+\vec{t}\cdot\vec{A})
=
\frac{1}{m}\left(I+\sum_{r=1}^{m^2-1}t_rA_r\right) .
}
Using the trace-log representation of the relative entropy (see Property \ref{DTL} in \rLmm{Drhoprop}), we have
\alg{
-m\frac{\partial}{\partial t_r}D(\pi\|\rho(\vec{t}))
=
\frac{\partial}{\partial t_r}{\rm Tr}\ln{(m\rho(\vec{t}))}
=\frac{1}{m}{\rm Tr}[A_r\rho(\vec{t})^{-1}].
}
Multiplying $A_r$ and taking the summation over $r$, we obtain
\alg{
-m\vec{A}\cdot{\rm grad}_{\vec{t}}D(\pi\|\rho(\vec{t}))
=
\sum_{r=1}^{m^2-1}\frac{1}{m}{\rm Tr}[A_r\rho(\vec{t})^{-1}]A_r 
=\rho(\vec{t})^{-1}-\frac{{\rm Tr}[\rho(\vec{t})^{-1}]}{m}I.
\laeq{patipati}
}
Due to the convexity of the relative entropy (see Property \ref{DCv} in \rLmm{Drhoprop}), the infimum in \req{houjicha} is achieved by the state $\rho(\vec{t})$ such that
$
{\rm grad}_{\vec t}D(\pi\|\rho(\vec{t}))
\propto
{\rm grad}_{\vec t}{\rm Tr}[W\rho(\vec{t})]
$,
which is equivalent to $\vec{A}\cdot{\rm grad}_{\vec{t}}D(\pi\|\rho(\vec{t}))\propto W$.
Thus, due to \req{patipati}, it follows that $\rho(\vec{t})^{-1}=\mu I+m\nu W$ for some $\mu,\nu\in\mbb{R}\backslash\{0\}$.
Multiplying $\rho(\vec{t})$ on both sides, and substituting the condition that ${\rm Tr}[W\rho(\vec{t})]=w$, we have that $\mu=m(1-w\nu)$.
From the normalization condition and the positive semidefiniteness of $\rho(\vec{t})$, we obtain \req{houjisha} and \req{houjijha}.

We consider the case where $w$ is sufficiently small so that $m|w|\ll\|W\|_2$.
Suppose now that the solution of \req{houjisha} and \req{houjijha} satisfies
\alg{
|\nu^*(w)|\cdot\|wI-W\|_\infty\ll1.
\laeq{spoon}
}
In this case, the expression \req{houjicha} is equivalent to
\alg{
\mathscr{D}_{\rm exp}(w;W,m)
=
-\frac{1}{m}\sum_{\ell=1}^\infty\frac{\nu^*(w)^{\ell}}{\ell}{\rm Tr}[(wI-W)^{\ell}] 
\laeq{tranpp}
}
and the condition \req{houjisha} is represented as
\alg{
\frac{1}{m}\sum_{\ell=0}^\infty \nu^{\ell}{\rm Tr}[(wI-W)^{\ell}]=1.
\laeq{tranp}
}
The expressions \req{tranp} and \req{tranpp} can be well approximated by the summations up to the order $\ell=2$.
Noting that we have $mw^2\ll{\rm Tr}[W^2]$ from the assumption, we obtain from \req{tranp} that
\alg{
\nu^*(w)
\approx
-\frac{mw}{{\rm Tr}[W^2]}.
\laeq{bakada}
}
The condition \req{spoon} is indeed satisfied, because we have
$
|\nu^*(w)|\cdot\|wI-W\|_\infty
\leq
2|\nu^*(w)|\cdot\|W\|_\infty
\leq
2|\nu^*(w)|\cdot\|W\|_2
\approx 2mw/\|W\|_2
\ll1
$.
Substituting \req{bakada} to \req{tranpp}, we obtain
\alg{
\mathscr{D}_{\rm exp}(w;W,m)
\approx
\frac{mw^2}{2{\rm Tr}[W^2]}.
}
This shows an improvement over the exponent that can be deduced from the results of \cite{popescu2005foundations,reimann2015generalization} by a constant factor (see \rApp{compPRPR} for the detail).\\

\begin{figure}[t]
\begin{center}
\includegraphics[bb={20 25 326 236}, scale=0.85]{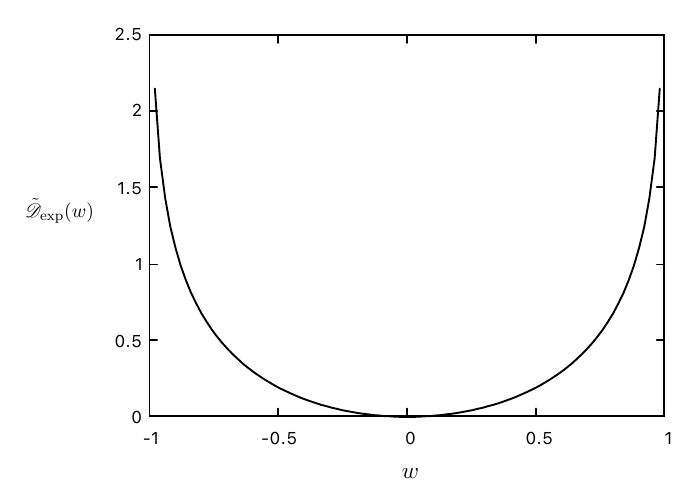}
\end{center}
\caption{
A graph of the function $\tilde{\mathscr{D}}_{\rm exp}(w):=\mathscr{D}_{\rm exp}(w;W=\proj{1}-\proj{3},m=3)$, which is given by \req{w3func}, is presented.
It goes to infinity in the limit of $w\rightarrow\pm1$.
}
\label{fig:w3}
\end{figure}

{\bf Example:}
We consider the case where $m=3$ and $W=\proj{1}-\proj{3}$ (see \rApp{prfnuw} for a detailed calculation).
The conditions \req{houjisha} and \req{houjijha} yield
\alg{
1-(w-k)\nu
\geq
0\;(k=-1,0,1)
}
and
\alg{
\frac{1}{3}\sum_{k=-1,0,1}\frac{1}{1-w\nu+k\nu}=1.
}
It follows that
\alg{
\nu^*(w)
=
\frac{1-3w^2-\sqrt{1+3w^2}}{3w(1-w^2)}.
\laeq{luzern}
}
The infimum of the relative entropy is calculated by \req{houjicha} to be
\alg{
\mathscr{D}_{\rm exp}(w;W=\proj{1}-\proj{3},m=3)
=
 \frac{1}{3}\sum_{k=-1,0,1} \ln{[1 - (w - k)\nu^* (w)]}.
 \laeq{w3func}
}
A graph of this function is depicted in \rFig{w3}.

\section{Conditional Concentration}
\lsec{condconc}

\rThm{LDP} implies that the probability of the set $\Omega\in\ca{S}(\ca{H}^S)$ is essentially the same as the probability of the state in $\Omega$ that is closest to the maximally mixed state in terms of the quantum relative entropy.
In this section, we prove that the total probability of the states in $\Omega$ that is $\epsilon$-away from the closest state $\rho^*$, conditioned by $\Omega$, is exponentially small with respect to $n$ for any $\epsilon>0$.
Thus, the total probability of $\Omega$ strongly concentrates around $\rho^*$.
We refer to this property as {\it conditional concentration}. 
 (As similar concepts, see e.g.~Section 11.6 in \cite{cover05} for the ``conditional limit theorem'' and \cite{la2015general} for the ``conditional large deviation property''.)
To be precise:

\bthm{condlim}
Let $\Omega\subseteq\ca{S}(\ca{H}^S)$ be such that ${\rm cl}({\rm int}\Omega)={\rm cl}\Omega$.
Define $\Omega_*\subseteq{\rm cl}\Omega$ by
\alg{
\Omega_*
:=
\left\{
\rho\in{\rm cl}\Omega
\left|
D(\pi\|\rho)
=
\min_{\rho'\in{\rm cl}\Omega}D(\pi\|\rho')
\right.\right\}.
}
Let $\ca{B}_\epsilon(\Omega_*)$ be the set of the $\epsilon$-neighbourhoods of $\Omega_*$, i.e.,
\alg{
\ca{B}_\epsilon(\Omega_*)
:=
\left\{\rho\in\ca{S}(\ca{H}^S)\left|\inf_{\sigma\in\Omega_*}\frac{1}{2}\|\rho-\sigma\|_1\leq\epsilon\right.\right\}.
}
For any $\epsilon>0$, there exists $\delta>0$ and it holds that
\alg{
-\limsup_{n\rightarrow\infty}
\frac{1}{n}
\ln{\frac{{\rm Pr}\left\{\left.\rho_\Psi\in\Omega\backslash\ca{B}_\epsilon(\Omega_*)\right|\Psi\sim\mu_n\right\}}{{\rm Pr}\left\{\left.\rho_\Psi\in\Omega\right|\Psi\sim\mu_n\right\}}}
\geq\delta.
\laeq{tanh}
}
\ethm

\bprf
Define nonnegative functions $f,g:{\rm cl}\Omega\rightarrow {\mathbb R}$ by
\alg{
f(\rho)=D(\pi\|\rho)-\min_{\rho'\in{\rm cl}\Omega}D(\pi\|\rho'),\quad
g(\rho)=\inf_{\sigma\in\Omega_*}\frac{1}{2}\|\rho-\sigma\|_1.
}
By definition, we have $f(\rho)\geq0$, $g(\rho)=0$ and
\alg{
f(\rho)=0 \;\Leftrightarrow\; g(\rho)=0.
\laeq{kiite}
}
Define
\alg{
\mathfrak{F}_\leq(\delta)
:=
\{\rho\in{\rm cl}\Omega|f(\rho)\leq\delta\}
\laeq{osaosa}
}
and
\begin{eqnarray}
\zeta(\delta):=\max_{\rho\in\mathfrak{F}_\leq(\delta)}g(\rho).
\label{def:zetapsi}
\end{eqnarray}
This is a monotonically nondecreasing function of $\delta$ by definition.
It satisfies $\lim_{\delta\rightarrow0}\zeta(\delta)=0$ as we prove later.
Thus, for any $\epsilon>0$, there exists $\delta>0$ such that
$
\mathfrak{F}_\leq(\delta)\subseteq\ca{B}_\epsilon(\Omega_*)
$.
It follows that
\alg{
{\rm Pr}\left\{\left.\rho_\Psi\in\Omega\backslash\ca{B}_\epsilon(\Omega_*)\right|\Psi\sim\mu_n\right\}
\leq
{\rm Pr}\left\{\left.\rho_\Psi\in\Omega\backslash\mathfrak{F}_\leq(\delta)\right|\Psi\sim\mu_n\right\}.
}
Hence, we have
\alg{
&
\limsup_{n\rightarrow\infty}
\frac{1}{n}
\ln{\frac{{\rm Pr}\left\{\left.\rho_\Psi\in\Omega\backslash\ca{B}_\epsilon(\Omega_*)\right|\Psi\sim\mu_n\right\}}{{\rm Pr}\left\{\left.\rho_\Psi\in\Omega\right|\Psi\sim\mu_n\right\}}}
\nn\\
&\leq
\limsup_{n\rightarrow\infty}
\frac{1}{n}
\ln{\frac{{\rm Pr}\left\{\left.\rho_\Psi\in\Omega\backslash\mathfrak{F}_\leq(\delta)\right|\Psi\sim\mu_n\right\}}{{\rm Pr}\left\{\left.\rho_\Psi\in\Omega\right|\Psi\sim\mu_n\right\}}}
\\
&=
\lim_{n\rightarrow\infty}
\frac{1}{n}
\ln{{\rm Pr}\left\{\left.\rho_\Psi\in\Omega\backslash\mathfrak{F}_\leq(\delta)\right|\Psi\sim\mu_n\right\}}
-
\lim_{n\rightarrow\infty}
\frac{1}{n}
\ln{{\rm Pr}\left\{\left.\rho_\Psi\in\Omega\right|\Psi\sim\mu_n\right\}}
\\
&=
-\inf_{\rho\in\Omega\backslash\mathfrak{F}_\leq(\delta)}D(\pi\|\rho)+\inf_{\rho\in\Omega}D(\pi\|\rho)
\laeq{pantpant}
\\
&=
-\inf_{\rho\in\Omega\backslash\mathfrak{F}_\leq(\delta)}D(\pi\|\rho)+\min_{\rho\in{\rm cl}\Omega}D(\pi\|\rho)
\laeq{pantpantt}
\\
&
\leq\delta,
\laeq{concop}
}
where \req{pantpant} follows from \rThm{LDP}, \req{pantpantt} from the continuity of $D(\pi\|\rho)$ with respect to $\rho$ and \req{concop} from \req{osaosa}.
For any sufficiently small $\delta>0$, \req{concop} is equal to $\delta$, which implies \req{tanh}.

We prove that $\zeta(\delta)$ defined by (\ref{def:zetapsi}) satisfies $\lim_{\delta\rightarrow0}\zeta(\delta)=0$.
The proof is along the same line as Appendix B-I of \cite{wakakuwa2017markovianizing}, which was based on an idea used in \cite{mixcomp1}. 
By definition, we have that
\alg{
\forall\delta>0,\:\exists\rho\in{\rm cl}\Omega\;:\;g(\rho)=\zeta(\delta)\wedge f(\rho)\leq\delta.
}
Define $\zeta_0:=\lim_{\delta\rightarrow0}\zeta(\delta)$. 
Due to the monotonicity, we have $\zeta(\delta)\geq\zeta_0$ for all $\delta>0$. 
Consequently, we have that
\begin{eqnarray}
\forall\delta>0,\:\exists\rho\in{\rm cl}\Omega\;:\;g(\rho)\geq\zeta_0\wedge f(\rho)\leq\delta.
\label{exsalpha}
\end{eqnarray}
Define ${\mathfrak F}_\geq(\zeta_0):=\{\rho\in{\rm cl}\Omega\:|\:g(\rho)\geq\zeta_0\}$. 
Due to the continuity of $g$, ${\mathfrak F}_\geq(\zeta_0)$ is a closed subset of ${\rm cl}\Omega$. 
Hence
$
\beta:=\min_{\rho\in{\mathfrak F}_\geq(\zeta_0)}f(\rho)
$
exists due to the continuity of $f$. 
By definition, we have that
\begin{eqnarray}
\forall\rho\in{\rm cl}\Omega:\:g(\rho)\geq\zeta_0\Rightarrow f(\rho)\geq\beta.
\label{exsbeta}
\end{eqnarray} 
Suppose now that $\zeta_0>0$. 
We have $f(\rho)>0$ for all $\rho\in{\mathfrak F}_\geq(\zeta_0)$ due to \req{kiite}. 
Thus we have $\beta>0$, in which case (\ref{exsbeta}) contradicts with (\ref{exsalpha}) because $\delta$ can be arbitrarily small.
\QED
\eprf

The conditional concentration property can be generalized to any continuous function of states as follows:

\bthm{condlimR}%{\bf[Conditional Concentration.]}
Let $v\in\mbb{N}$ and let $\gamma:\ca{S}(\ca{H}^S)\rightarrow\mbb{R}^v$ be a uniformly continuous function.
Define a function $D_\gamma:\mbb{R}^v\rightarrow\mbb{R}$ by
\alg{
D_\gamma(x)
=
\inf_{\rho\in\gamma^{-1}(\{x\})}
D(\pi\|\rho).
}
Let $\Gamma\subseteq\mbb{R}^v$ be an open set and define $\Gamma_*\subset{\rm cl}\Gamma$ by
\alg{
\Gamma_*
:=
\left\{
x\in{\rm cl}\Gamma
\left|
D_\gamma(x)
=
\min_{x'\in{\rm cl}\Gamma}D_\gamma(x')
\right.\right\}.
}
Furthermore, let $\mfk{B}_\iota(\Gamma_*)$ be the set of the $\iota$-neighbourhoods of $\Gamma_*$, i.e.,
\alg{
\ca{B}_\iota(\Gamma_*)
:=
\left\{x\in\mbb{R}^v\left|\min_{x_*\in\Gamma_*}\|x-x_*\|\leq\iota\right.\right\}.
}
For any $\iota>0$, there exists $\delta>0$ and it holds that
\alg{
-\limsup_{n\rightarrow\infty}
\frac{1}{n}
\ln{\frac{{\rm Pr}\left\{\left.\gamma(\rho_\Psi)\in\Gamma\backslash\mathfrak{B}_\iota(\Gamma_*)\right|\Psi\sim\mu_n\right\}}{{\rm Pr}\left\{\left.\gamma(\rho_\Psi)\in\Gamma\right|\Psi\sim\mu_n\right\}}}
\geq\delta.
\laeq{tanhR}
}
\ethm

\bprf
Define $\Omega:=\gamma^{-1}(\Gamma)$.
Since $\gamma$ is a continuous function, $\Omega$ is an open set and thus ${\rm cl}({\rm int}\Omega)={\rm cl}\Omega$.
Define $\Omega_*\subset\ca{S}(\ca{H}^S)$ and $\ca{B}_\epsilon(\Omega_*)\subset\ca{S}(\ca{H}^S)$ in the same way as in \rThm{condlim}.
Due to the uniform continuity of $\gamma$, for any $\iota>0$, there exists $\epsilon>0$ such that $\gamma(\ca{B}_\epsilon(\Omega_*)\cap\Omega)\subseteq\mathfrak{B}_\iota(\Gamma_*)$.
Thus, $\gamma(\rho_\Psi)\in\Gamma\backslash\mathfrak{B}_\iota(\Gamma_*)$ implies $\gamma(\rho_\Psi)\in\Gamma\backslash\gamma(\ca{B}_\epsilon(\Omega_*)\cap\Omega)$, which in turn implies $\rho_\Psi\in\Omega\backslash\ca{B}_\epsilon(\Omega_*)$. 
Consequently, we have
\alg{
{\rm Pr}\left\{\left.\gamma(\rho_\Psi)\in\Gamma\backslash\mathfrak{B}_\iota(\Gamma_*)\right|\Psi\sim\mu_n\right\}
\leq
{\rm Pr}\left\{\left.\rho_\Psi\in\Omega\backslash\ca{B}_\epsilon(\Omega_*)\right|\Psi\sim\mu_n\right\}.
}
By definition, we have
\alg{
{\rm Pr}\left\{\left.\gamma(\rho_\Psi)\in\Gamma\right|\Psi\sim\mu_n\right\}
=
{\rm Pr}\left\{\left.\rho_\Psi\in\Omega\right|\Psi\sim\mu_n\right\}.
}
Thus, from \rThm{condlim}, we obtain
\alg{
&
-\limsup_{n\rightarrow\infty}
\frac{1}{n}
\ln{\frac{{\rm Pr}\left\{\left.\gamma(\rho_\Psi)\in\Gamma\backslash\mathfrak{B}_\iota(\Gamma_*)\right|\Psi\sim\mu_n\right\}}{{\rm Pr}\left\{\left.\gamma(\rho_\Psi)\in\Gamma\right|\Psi\sim\mu_n\right\}}}
\nn\\
&
\geq
-\limsup_{n\rightarrow\infty}
\frac{1}{n}
\ln{\frac{{\rm Pr}\left\{\left.\rho_\Psi\in\Omega\backslash\ca{B}_\epsilon(\Omega_*)\right|\Psi\sim\mu_n\right\}}{{\rm Pr}\left\{\left.\rho_\Psi\in\Omega\right|\Psi\sim\mu_n\right\}}}
\geq\delta
}
and complete the proof.
\QED
\eprf

\section{Coherence of Random Pure States}
\lsec{coherence}

We apply approachs presented in the previous sections to investigate the large deviation property of coherence of random pure states in a {\it single} quantum system.
Ref.~\cite{singh2016average} proved that coherence of random pure states in a system with large dimension strongly concentrates around a near-maximal value.
To be precise, let $\ca{H}_n$ be a $n$-dimensional Hilbert space with a fixed orthonormal basis $\{|e_l\rangle\}_{l=1}^n$ for each $n\in\mbb{N}$.
For each $\ket{\Psi}\in\ca{H}_n$, the relative entropy of coherence \cite{baumgratz14} is calculated to be
\alg{
C_r(\Psi)
:=
H(\{p_\Psi(l)\}_{l=1}^n)
=
-\sum_{l=1}^{n}p_\Psi(l)\ln{p_\Psi(l)},
\laeq{dfnShc}
}
where $p_\Psi(l):=|\langle e_l|\Psi\rangle|^2$.
It was proved in Ref.~\cite{singh2016average} (see Theorem 1 therein) that, for any $n\geq3$ and $\omega>0$, it holds that
\alg{
{\rm Pr}\left\{\left.\left|\frac{C_r(\Psi)}{\ln{n}}-\frac{H_n-1}{\ln{n}}\right|>\omega\right|\Psi\sim\nu_n\right\}
\leq
2\exp\left(-\frac{n\omega^2}{36\pi^3\ln{2}}\right),
\laeq{monimo}
}
where $\nu_n$ is the unitarily invariant measure on $\ca{H}_n$ and $H_n=\sum_{k=1}^n(1/k)$.
The proof in Ref.~\cite{singh2016average} is based on evaluation of the Lipschitz constant of $C_r$ and application of Levy's lemma, which followed the lines of Section III in \cite{hayden2006aspects} to prove concentration of entanglement.
Noting that $\lim_{n\rightarrow\infty}(H_n-1)/\ln{n}=1$,
we obtain an asymptotic bound from \req{monimo} as
\alg{
-\limsup_{n\rightarrow\infty}
\frac{1}{n}\ln{\rm Pr}\left\{\left.1-\frac{C_r(\Psi)}{\ln{n}}>\omega\right|\Psi\sim\nu_n\right\}
\geq
\frac{\omega^2}{36\pi^3\ln{2}},
\laeq{hokh}
}
which provides a lower bound on the exponent of coherence of random pure states.
In the following, we provide a large deviation bound that is complementary to \req{hokh}, namely an upper bound on the exponent:

\bthm{coherenceLD}
For any $0<\omega<1$,
it holds that
\alg{
-\liminf_{n\rightarrow\infty}\frac{1}{n}\ln{{\rm Pr}\left\{\left.1-\frac{C_r(\Psi)}{\ln{n}}\geq\omega\right|\Psi\sim\nu_n\right\}}
\leq
\ln{\left(\frac{1}{1-\omega}\right)}.
\laeq{hoch}
}
\ethm

\bprf
Fix arbitrary $\delta$ satisfying $0<\delta<1-\omega$ and define $p^*(\Psi):=\max_{1\leq l\leq n}{p_\Psi(l)}$.
A simple calculation yields
\alg{
H(\{p_\Psi(l)\}_{l=1}^n)
\leq
h(p^*(\Psi))+(1-p^*(\Psi))\ln{n}
\leq
1+(1-p^*(\Psi))\ln{n},
}
where $h$ is the binary entropy defined by $h(x)=-x\ln{x}-(1-x)\ln{(1-x)}$.
Thus, for any $n\geq e^{1/\delta}$, we have
\alg{
\frac{C_r(\{\Psi\})}{\ln{n}}
\leq
\delta+(1-p^*(\Psi)),
}
which implies
\alg{
{\rm Pr}\left\{\left.1-\frac{C_r(\Psi)}{\ln{n}}\geq\omega\right|\Psi\sim\nu_n\right\}
\geq
{\rm Pr}\left\{\left.p^*(\Psi)\geq\omega+\delta\right|\Psi\sim\nu_n\right\}.
}
It follows that
\alg{
-\liminf_{n\rightarrow\infty}\frac{1}{n}\ln{\rm Pr}\left\{\left.1-\frac{C_r(\Psi)}{\ln{n}}\geq\omega\right|\Psi\sim\nu_n\right\}
\leq
-\liminf_{n\rightarrow\infty}\frac{1}{n}\ln{\rm Pr}\left\{\left.p^*(\Psi)\geq\omega+\delta\right|\Psi\sim\nu_n\right\}.
}
As we prove below, we have
\alg{
  -\lim_{n\rightarrow\infty}\frac{1}{n}\ln{\rm Pr}\left\{\left.p^*(\Psi)\geq\kappa\right|\Psi\sim\nu_n\right\}
=\ln{\left(\frac{1}{1-\kappa}\right)}.  
\laeq{bruk}
}
for any $0<\kappa<1$.
Substituting $\omega+\delta$ to $\kappa$ and taking the limit of $\delta\rightarrow0$, we arrive at \req{hoch}.

To prove \req{bruk}, note that the condition $p^*(\Psi)\geq\kappa$ is equivalent to the condition that there exists $l\:(1\leq l\leq n)$ satisfying $p_\Psi(l)\geq \kappa$.
Thus, we have that
\alg{
{\rm Pr}\left\{\left.|\langle e_1|\Psi\rangle|^2\geq \kappa\right|\Psi\sim\nu_n\right\}
&\leq
{\rm Pr}\left\{\left.p^*(\Psi)\geq\kappa\right|\Psi\sim\nu_n\right\}
\leq
\sum_{l=1}^n{\rm Pr}\left\{\left.|\langle e_l|\Psi\rangle|^2\geq \kappa\right|\Psi\sim\nu_n\right\}.
\laeq{yuta}
}
Define a normalized state vector
$
|+_n\rangle
:=
(1/\sqrt{n})\sum_{l=1}^n|e_l\rangle
$.
Due to the unitary invariance of $\nu_n$, we have
\alg{
{\rm Pr}\left\{\left.|\langle e_l|\Psi\rangle|^2\geq \kappa \right|\Psi\sim\nu_n\right\}
=
{\rm Pr}\left\{\left.|\langle+_n|\Psi\rangle|^2\geq \kappa \right|\Psi\sim\nu_n\right\}
}
for any $l$.
It follows from \req{yuta} that
\alg{
{\rm Pr}\left\{\left.|\langle+_n|\Psi\rangle|^2\geq \kappa \right|\Psi\sim\nu_n\right\}
\leq
{\rm Pr}\left\{\left.p^*(\Psi)\geq\kappa\right|\Psi\sim\nu_n\right\}
\leq
n{\rm Pr}\left\{\left.|\langle+_n|\Psi\rangle|^2\geq \kappa \right|\Psi\sim\nu_n\right\}.
}
Hence, to prove \req{bruk}, it suffices to prove that
\alg{
-\lim_{n\rightarrow\infty}\frac{1}{n}\ln{{\rm Pr}\left\{\left.|\langle+_n|\Psi\rangle|^2\geq\kappa \right|\Psi\sim\nu_n\right\}}
=\ln{\left(\frac{1}{1-\kappa}\right)}
\laeq{tashikani}
}
for any $0<\kappa<1$.

To prove \req{tashikani}, let $Z=(X,Y)^T$ be a $2$-dimensional real vector random variable that obeys the standard normal distribution (SND), i.e., $Z$ obeys a probability distribution $G_2(Z)=G_1(X)G_1(Y)$, where $G_1$ is defined by \req{dfnSND}.
Let $Z_1,\cdots,Z_l,\cdots,Z_n$ be an i.i.d. sequence of the random variable $Z$, where $Z_l\equiv(X_l,Y_l)^T$.
Define a random unnormalized vector $\ket{\tilde{\Psi}}$ by
\alg{
\ket{\tilde{\Psi}}
=
\sum_{1\leq l\leq n}(X_{l}+iY_{l})\ket{e_{l}},
\laeq{tadori}
}
and a normalized one by $|\bar{\Psi}\rangle\equiv\ket{\tilde{\Psi}}/\|\ket{\tilde{\Psi}}\|$.
With a slight abuse of notation, we denote $\bar{\Psi}\sim{\rm SND}(n)$.
Due to \rLmm{unitaryinvariantmeasure} in \rSec{parsii}, the random state $\ket{\Psi}$ distributed with respect to the unitarily invariant measure on $\ca{H}_n$ is equivalent to the random state $|\bar{\Psi}\rangle$.
Thus, the probability in the L.H.S.~of \req{tashikani} is equal to 
$
{\rm Pr}\left\{\left.|\langle+_n|\bar{\Psi}\rangle|^2\geq\kappa \right|\Psi\sim\nu_n\right\}
$.

Define $\Xi\subset\mbb{R}^3$ and $\Theta(\kappa)\subset\Xi$ for $0<\kappa<1$ by
\alg{
\Xi
&:=
\left\{(s,t)\in\mbb{R}\times\mbb{R}^2\left|s>0,\frac{\|t\|^2}{s}<1\right.\right\},
\laeq{DEFX}
\\
\Theta(\kappa)
&:=
\left\{(s,t)\in\Xi\left|
\kappa\leq\frac{\|t\|^2}{s}\right.\right\}.
}
Define functionals $\eta:\Lambda(\mbb{R}^{2})\rightarrow\mbb{R}$ and $\zeta:\Lambda(\mbb{R}^{2})\rightarrow\mbb{R}^2$ by
\alg{
\eta(P)
:=
\int_{z\in\mbb{R}^2} \|z\|^2P(z)dz,
\quad
\zeta(P)
:=
\int_{z\in\mbb{R}^2} zP(z) dz,
}
in addition to $\xi:\Lambda(\mbb{R}^{2})\rightarrow\mbb{R}$ and $\Phi:\Lambda(\mbb{R}^{2})\rightarrow\mbb{R}^3$ by
\alg{
\xi(P)
:=
\frac{\|\zeta(P)\|^2}{\eta(P)},
\quad
\Phi(P)
:=
(\eta(P),\zeta(P)).
}
%The condition $\xi(P)\geq\kappa$ is equivalent to $\Phi(P)\in\Theta(\kappa)$.
A simple calculation yields
\alg{
|\langle+_n|\bar{\Psi}\rangle|^2
=
\frac{\left\|\frac{1}{n}\sum_{l=1}^nZ_l\right\|^2}{\frac{1}{n}\sum_{1\leq l\leq n}
  \|Z_l\|^2}=\xi(\hat{P}_n^{Z^n}), 
 \laeq{komamaku}
}
where $\hat{P}_n^{Z^n}$ denotes the empirical distribution of the sequence.
Thus, the condition $|\langle +_n|\bar{\Psi}\rangle|^2\geq\kappa$ is equivalent to the condition that $\xi(\hat{P}_n^{Z^n})\geq\kappa$, which in turn is equivalent to
$
\Phi(\hat{P}_n^{Z^n})
\in
\Theta(\kappa)
$.
Hence, it suffices to prove that
\alg{
-\lim_{n\rightarrow\infty}
\frac{1}{n}\ln{{\rm Pr}\left\{\hat{P}_n^{Z^n}
\in\Phi^{-1}(\Theta(\kappa))\right\}}
=\ln{\left(\frac{1}{1-\kappa}\right)}
\laeq{perieerri}
}
for any $0<\kappa<1$.

To prove \req{perieerri}, define a function $D^*:\mbb{R}\times\mbb{R}^2\rightarrow\mbb{R}$ by
\alg{
D^*(s,t):=D(\Phi^{-1}\{(s,t)\}\|G_2).
}
As we prove below, it holds that
\alg{
D^*(s,t)
=
\begin{cases}
\frac{s}{2}-\ln{\left(\frac{s-\|t\|^2}{2}\right)}-1&(s,t)\in\Xi,
\laeq{smetana}
\\
+\infty&(s,t)\notin\Xi.
\end{cases}
}
It is obvious that $D^*(s,t)$ is continuous in $\Xi$, and that ${\rm int}\Xi=\Xi$ from \req{DEFX}.
Thus, we can apply \rThm{sanov}.
A simple calculation yields
\alg{
\inf_{(s,t)\in{\rm cl}\Theta(\kappa)} D^*(s,t)
=
\inf_{(s,t)\in{\rm int}\Theta(\kappa)} D^*(s,t)
=
-\ln{(1-\kappa)},
}
which leads to \req{perieerri}.

To prove \req{smetana}, note that the condition $\Phi(P)=(s,t)$ is equivalent to
\alg{
\int_{z\in\mbb{R}^2}  \|z\|^2P(z)dz
=
s
\wedge
\int_{z\in\mbb{R}^2} zP(z)dz
=t. 
}
For $(s,t)\in\Xi$, suppose that there exists $P\in\Lambda(\mbb{R}^{2m})$ such that
\alg{
P\in\Phi^{-1}(s,t),\;D(P\|Q)<\infty,\;P(z)>0\:(\forall z\in\mbb{R}^2).
\laeq{marumoru}
}
From \rThm{csiszar}, it follows that
\alg{
D^*(s,t)
&
=
\max_{\mu\in\mbb{R},\nu\in\mbb{R}^2}\left[-\ln{\left\{\int_{z\in\mbb{R}^2}{ G_2(z)e^{-\mu (\|z\|^2-s)-\nu\cdot(z-t)}dz}\right\}}\right]
\\
&=
-\min_{\mu\in\mbb{R},\nu\in\mbb{R}^2}\left[\mu s + \nu \cdot  t + \frac{\|\nu\|^2}{2(1 + 2\mu)} - \ln{(1 + 2\mu)}\right],
\laeq{renslayer}
}
where the second line follows by taking the Gaussian integral.
The minimum is achieved by 
\alg{
(\mu^*,\nu^*)=
\left(\frac{1}{s-\|t\|^2}-\frac{1}{2}\:,\:\frac{-2t}{s-\|t\|^2}\right)
}
which yields
\alg{
D^*(s,t)
=
\frac{s}{2}-\ln{\left(\frac{s-\|t\|^2}{2}\right)}-1.
}
It is straightforward to verify that a probability distribution $P^*$ defined by 
\alg{
P^*(z)
=
\frac{G_2(z)e^{-\mu^* (\|z\|^2-s)-\nu^*(z-t)}dz}{\int_{z'\in\mbb{R}^2}{ G_2(z')e^{-\mu^* (\|z'\|^2-s)-\nu^*(z'-t)}dz'}}
}
indeed satisfies the condition \req{marumoru}.
For $(s,t)\notin{\rm cl}\Xi$, we have $\Phi^{-1}(s,t)=\emptyset$ because $\eta(P)\geq0$ and $\xi(P)\leq1$ for any $P\in\Lambda(\mbb{R}^{2m})$.
For $(s,t)\in\partial{\Xi}$, any $P\in\Phi^{-1}(s,t)$ satisfies $\eta(P)=\|\zeta(P)\|^2$.
Thus, there exists $\iota_{(s,t)}\in\mbb{R}^2$ such that $P(z)=\delta(z-\iota_{(s,t)})$, which implies $D(P\|G_2)=+\infty$.
In both cases, we have $D^*(s,t)=+\infty$.
All in all, we obtain \req{perieerri} and complete the proof.
\QED
\eprf

Noting that $\ln{(\frac{1}{1-\omega})}=\sum_{\ell=1}^\infty\ell^{-1}\omega^{\ell}$, there remains a quadratic gap between the lower bound \req{hokh} and the upper bound \req{hoch} with respect to $\omega$, in addition to the constant factor $36\pi^2\ln{2}$. 
To close this gap and obtain the exact exponent for the coherence of random pure states is left as a future work.

\section{Conclusion} 
\lsec{conclusion}

We have investigated the properties of the random quantum states induced from the uniformly random pure states on a bipartite system by taking the partial trace over the larger subsystem.
We have proved that the probability of the induced random state within a given set decreases no slower or faster than exponential in the dimension of the system traced out.
The exponent is given by the quantum relative entropy between the maximally mixed state and the set of states to be considered.
We have also shown that the total probability of a given set strongly concentrates around the element closest to the maximally mixed state.
The same method was applied to investigating the large deviation property of coherence of random quantum states in a single system.

The main result (\rThm{LDP}) can be regarded as a ``Sanov-type'' theorem for atypicality of random quantum states. 
Quantum extensions of Sanov’s theorem have been proposed in the context of quantum hypothesis testing \cite{bjelakovic2005quantum,audenaert2008asymptotic,notzel2014hypothesis}.
However, it was pointed out in \cite{nagaoka2007information} (see Remark 8 therein) that finding a meaningful and useful quantum extension of Sanov’s theorem remains open. 
\rThm{LDP} provides some insights on this problem from the viewpoint of random quantum states.
However, it cannot be regarded exactly as a quantum version of Sanov's theorem.
Note that, in Eq.~\req{gojure}, the first argument in the relative entropy is fixed to the maximally mixed state, and the infimum is taken over the second argument thereof, as opposed to the original Sanov’s theorem and those presented in \cite{bjelakovic2005quantum,audenaert2008asymptotic,notzel2014hypothesis}.

We conclude this paper with a list of problems that are left for future work.

As discussed in the introduction, the large deviation bound obtained in this paper is valid only in the asymptotic limit, where the dimension of the subsystem traced out goes to infinity. 
It tells us nothing in the case where the dimension is finite. 
To obtain a large deviation bound that is valid in the finite-dimensional case and is tight in the asymptotic limit is left as an important open problem.

In the main result of this paper, we assumed that the pure state is chosen randomly from the entire Hilbert space, which is represented as a tensor product of two Hilbert spaces. 
The concept of measure concentration has also been applied to investigating properties of random pure quantum states in a {\it single} system \cite{reimann2015generalization,singh2016average} or in a {\it subspace} of the tensor product of two Hilbert spaces \cite{hayden2006aspects,popescu2006entanglement,popescu2005foundations}. 
It would be fruitful to explore the applicability of the method presented in this paper to those cases, besides the analysis of the coherence property of random pure states presented in \rSec{coherence}.

The Haar measure on the set of the unitary operators is generated from the multivariate normal distribution, in the same way as the unitary invariant measure on the set of the normalized pure states is (see e.g.~Section 7.2 in \cite{watrous2018theory}). Hence, it would be natural to expect that the approach presented in this paper is extended to investigating the large deviation property of random unitary operators \cite{hiai2000large}.

Ref.~\cite{low2009large} investigated the concentration bounds obtained in \cite{hayden2006aspects} from the viewpoint of the unitary designs and the state designs. 
It proved that a large deviation bound, similar to Eq.~\req{kanna}, holds when the state $t$-designs or the unitary $t$-designs are used instead of the unitary invariant measure or the Haar measure. 
In this direction, it would be natural to consider an extension of \rThm{LDP} to the case where the unitary invariant measure is replaced by the state $t$-designs.

Finally, it would be nice to apply the main result (\rThm{LDP}) to concrete examples besides those presented in \rSec{formresults} and \rSec{examples}. One interesting scenario would be the case where the system $S$ is composed of two subsystems. Applying \rThm{LDP}, we may investigate the large deviation property on the probability of the random state induced on $S$ being entangled.
This scenario was addressed in Section V.B of \cite{hayden2006aspects}.

%Finally, let us mention the potential applicability of the results of this paper to the foundation of non-equilibrium quantum statistical mechanics. @@@

{\it Note added.---} After the completion of this work, we became aware that Eq.~\req{gojure} and Eq.~\req{tashikani} can be proved much more simply from the result of \cite{zyczkowski2001induced} (see Eqs.~(3.6) and (2.3) therein).

\section*{Acknowledgement}

EW acknowledges support from MEXT Quantum Leap Flagship Program (MEXT QLEAP) Grant No. JPMXS0120319794.

%\begin{acknowledgments}
%\end{acknowledgments}
%
%\hfill
%

\appendix

\section{Proof of \rThm{sanov}}
\lapp{prfSanovCont}

We provide proof of  \rThm{sanov}.
First, we consider the case where $\Theta$ is convex and ${\rm cl}({\rm int}\Theta)={\rm cl}\Theta$.
Second, we extend the result to the case where these conditions are not necessarily satisfied.

\subsection{Proof for Convex Regular Subsets}

Assume that $\Theta$ is convex and ${\rm cl}({\rm int}\Theta)={\rm cl}\Theta$.
We prove that
\alg{
-\lim_{n\rightarrow\infty}\frac{1}{n}\ln{{\rm Pr}\{\bm{F}(\hat{P}_n^{Z^n})\in \Theta\}}
=
\inf_{\bm{t}\in\Theta}\mfk{D}(\bm{t}),
\laeq{jyaruj}
}
which implies \req{smetana1} and \req{smetana2}.

Define $B_m:=[-m,m)^{\times v}$, $\Lambda_m:=\{P\in\Lambda|P(B_m)=1\}$ for each $m\in\mbb{N}$ and let $\Lambda^*:=\cup_{m=1}^\infty\Lambda_m$.
It is obvious that $\lim_{m\rightarrow\infty}Q(B_m)=1$ and that the function $\bm{F}|_{\Lambda_m}$ is linear.
Due to \rThm{orig}, it suffices to prove that 
\alg{
\inf_{P\in \bm{F}^{-1}(\Theta)\cap\Lambda^*}D(P\|Q)
=
\inf_{\bm{t}\in \Theta}\mfk{D}(\bm{t})
\laeq{kaijs}
}
and that $\bm{F}$ is $\tau$-continuous in $\Lambda_m$ for each $m\in\mbb{N}$.

To prove \req{kaijs}, fix an arbitrary $\epsilon,\delta>0$ and define an open convex set
\alg{
\ca{I}^\epsilon(\Theta)
:=
\left\{\bm{t}\in \Theta\left|\inf_{\bm{\tau}\in\partial\Theta}\|\bm{t}-\bm{\tau}\|>\epsilon\right.\right\}.
}
For any $P\in\bm{F}^{-1}(\ca{I}^\epsilon(\Theta))$ and $m\in\mbb{N}$, define $P_m\in\Lambda_m$ by
\alg{
P_m(z)
=
\begin{cases}
\frac{P(z)}{P(B_m)}&(z\in B_m)\\
0&(z\notin B_m)
\end{cases}.
}
We have
\alg{
\bm{F}(P_m)
&=\frac{1}{P(B_m)}\int_{z\in B_m}P(z)\bm{f}(z)dz,\\
D(P_m\|Q)
&=
\frac{1}{P(B_m)}\int_{z\in B_m}P(z)\ln\frac{P(z)}{Q(z)}dz
-
\ln P(B_m)
}
and $\lim_{m\rightarrow\infty}P(B_m)=1$.
Thus, for any sufficiently large $m$, it holds that
\alg{
\|\bm{F}(P_m)-\bm{F}(P)\|&<\epsilon,
\laeq{surel}
\\
D(P_m\|Q)-D(P\|Q)&\leq\delta.
\laeq{burel}
}
It follows from \req{surel} that $P_m\in\bm{F}^{-1}(\Theta)\cap\Lambda^*$.
Thus, from \req{burel}, we have
\alg{
\inf_{P\in \bm{F}^{-1}(\ca{I}^\epsilon(\Theta))}D(P\|Q)
\geq
\inf_{P\in \bm{F}^{-1}(\Theta)\cap\Lambda^*}D(P\|Q)-\delta.
\laeq{furi}
}
We take the limit of $\epsilon,\delta\rightarrow0$.
Due to the continuity of $\mfk{D}(\bm{t})$ and the assumption that ${\rm cl}({\rm int}\Theta)={\rm cl}\Theta$, the L.H.S.~of \req{furi} is evaluated as
\alg{
\lim_{\epsilon\rightarrow0}\inf_{P\in \bm{F}^{-1}(\ca{I}^\epsilon(\Theta))}D(P\|Q)
=
\lim_{\epsilon\rightarrow0}\inf_{\bm{t}\in \ca{I}^\epsilon(\Theta)}\mfk{D}(\bm{t})
=
\inf_{\bm{t}\in \Theta}\mfk{D}(\bm{t}).
}
Thus, we arrive at
\alg{
\inf_{\bm{t}\in \Theta}\mfk{D}(\bm{t})
\geq
\inf_{P\in \bm{F}^{-1}(\Theta)\cap\Lambda^*}D(P\|Q).
}
The inverse inequality is straightforward as
\alg{
\inf_{\bm{t}\in \Theta}\mfk{D}(\bm{t})
=\inf_{P\in \bm{F}^{-1}(\Theta)}D(P\|Q)
\leq
\inf_{P\in \bm{F}^{-1}(\Theta)\cap\Lambda^*}D(P\|Q).
}
Thus, we obtain \req{kaijs}.

To prove $\tau$-continuity of $\bm{F}$ in $\Lambda_m$, we first assume that $w=1$.
For $l\in\mbb{N}$ and $\mu\equiv(\mu_1,\cdots,\mu_v)\in\mbb{Z}^{\times v}$,
define $\mbb{R}_{\mu,l,m}^v\subseteq\mbb{R}^v$ by
\alg{
\mbb{R}_{\mu,l,m}^v
:=
\bigtimes_{s=1}^v\left[\frac{m\mu_s}{l},\frac{m(\mu_s+1)}{l}\right).
}
It is obvious that $\{\mbb{R}_{\mu,l,m}^v\}_{\mu\in\{-l,\cdots,0,\cdots,l-1\}^{\times v}}$ is a partition of $B_m$.
For any $P,Q\in\Lambda_m$, it holds that
\alg{
\int_{z\in\mbb{R}_{\mu,l,m}^v}          P(z)f(z)dz
&\leq
\sup_{z\in\mbb{R}_{\mu,l,m}^v}      f(z)\cdot\int_{z\in\mbb{R}_{\mu,l,m}^v}           P(z)dz,
\\
\int_{z\in\mbb{R}_{\mu,l,m}^v}          Q(z)f(z)dz
&\geq
\inf_{z\in\mbb{R}_{\mu,l,m}^v}      f(z)\cdot\int_{z\in\mbb{R}_{\mu,l,m}^v}           Q(z)dz.
}
Thus,  we have
\alg{
&\int_{z\in\mbb{R}_{\mu,l,m}^v}(P(z)-Q(z))f(z)dz
\nn\\
&\leq
\sup_{z\in\mbb{R}_{\mu,l,m}^v}  \!\!\!    f(z)\cdot\int_{z'\in\mbb{R}_{\mu,l,m}^v}    \!\!\!  \!\!\!     (P(z')-Q(z'))dz'
+
\left(\sup_{z\in\mbb{R}_{\mu,l,m}^v}  \!\!\!    f(z)-\inf_{z\in\mbb{R}_{\mu,l,m}^v} \!\!\!     f(z)\right) \cdot \int_{z'\in\mbb{R}_{\mu,l,m}^v}     \!\!\!  \!\!\!  Q(z')dz'
\\
&\leq
\delta\cdot f_m^*
+
\Delta_{f,l,m}^*\cdot Q(\mbb{R}_{\mu,l,m}^v),
}
where we defined 
\alg{
\delta:=\max_{\mu}|Q(\mbb{R}_{\mu,l,m}^v)-P(\mbb{R}_{\mu,l,m}^v)|,
\quad
\Delta_{f,l,m}^*
:=
\max_{\mu}
\left(\sup_{z\in\mbb{R}_{\mu,l,m}^v}      f(z)-\inf_{z\in\mbb{R}_{\mu,l,m}^v}      f(z)\right)
 \laeq{sorea}
}
and $f_m^*:=\sup_{z\in B_m}f(z)$.
The maximizations in \req{sorea} are taken over all $\mu\in\{-l,\cdots,l-1\}^{\times v}$.
It follows that
\alg{
\bm{F}(P)-\bm{F}(Q)
&=
\int_{z\in[-m,m)^{\times v}}(P(z)-Q(z))f(z)dz
\\
&=
\sum_{\mu}\int_{z\in\mbb{R}_{\mu,l,m}^v}(P(z)-Q(z))f(z)dz
\\
&\leq
(2l)^v\delta\cdot f_m^*
+
\Delta_{f,l,m}^*\sum_{\mu}Q(\mbb{R}_{\mu,l,m}^v)
\\
&=
(2l)^v\delta\cdot f_m^*
+
\Delta_{f,l,m}^*.
}
Exchanging the roles of $P$ and $Q$, we also have
\alg{
\bm{F}(Q)-\bm{F}(P)
\leq
(2l)^v\delta\cdot f_m^*
+
\Delta_{f,l,m}^*.
}
Thus, we arrive at
\alg{
|\bm{F}(P)-\bm{F}(Q)|
\leq
(2l)^v\delta\cdot f_m^*
+
\Delta_{f,l,m}^*.
}
Due to the continuity of $f$, we have $f_m^*<\infty$ and $\lim_{l\rightarrow\infty}\Delta_{f,l,m}^*=0$.
Thus, we conclude that $\bm{F}$ is $\tau$-continuous in $\Lambda_m$.
It is straightforward to extend the above argument to the case where $w\geq2$, which completes the proof.
\QED

\subsection{Extension to General Subsets}

It remains to consider the case where $\Theta$ is not necessarily convex and ${\rm cl}({\rm int}\Theta)={\rm cl}\Theta$ does not necessarily hold.
For $l\in\mbb{N}$ and $\mu\equiv(\mu_1,\cdots,\mu_w)\in\mbb{Z}^{\times w}$,
define $\mbb{R}_{\mu,l}^w\subseteq\mbb{R}^w$ by
\alg{
\mbb{R}_{\mu,l}^w
:=
\bigtimes_{r=1}^w\left[\frac{\mu_r}{l},\frac{\mu_r+1}{l}\right).
}
Any $\mbb{R}_{\mu,l}^w$ is a convex set and satisfies ${\rm cl}({\rm int}\mbb{R}_{\mu,l}^w)={\rm cl}\mbb{R}_{\mu,l}^w$.
Thus, from \req{jyaruj}, it holds that
\alg{
-\lim_{n\rightarrow\infty}\frac{1}{n} \ln{{\rm Pr}\{\bm{F}(\hat{P}_n^{Z^n}) \in  \mbb{R}_{\mu,l}^w\}}
=  
\inf_{\bm{t}\in\mbb{R}_{\mu,l}^w}\mfk{D}(\bm{t}).
\laeq{yakis}
}
Define
\alg{
\mbb{Z}_{\Theta,+}^{w,l}
:=
\{\mu\in\mbb{Z}^{w}:\mbb{R}_{\mu,l}^w\cap\Theta\neq\emptyset\},
\quad
\mbb{Z}_{\Theta,-}^{w,l}
:=
\{\mu\in\mbb{Z}^{w}:\mbb{R}_{\mu,l}^w\subset\Theta\}
}
and
\alg{
\Theta_+^l
:=
\bigcup_{\mu\in\mbb{Z}_{\Theta,+}^{w,l}}\mbb{R}_{\mu,l}^w,
\quad
\Theta_-^l
:=
\bigcup_{\mbb{Z}_{\Theta,-}^{w,l}}\mbb{R}_{\mu,l}^w.
\laeq{eaeaso}
}
It is obvious that
\alg{
{\rm Pr}\{\bm{F}(\hat{P}_n^{Z^n})\in \Theta_\pm^l\}
=
\sum_{\mu\in\mbb{Z}_{\Theta,\pm}^{w,l}}    {\rm Pr}\{\bm{F}(\hat{P}_n^{Z^n})\in\mbb{R}_{\mu,l}^w\}.
}
Hence, we have
\alg{
\max_{\mu\in\mbb{Z}_{\Theta,\pm}^{w,l}}  {\rm Pr}\{\bm{F}(\hat{P}_n^{Z^n})\in\mbb{R}_{\mu,l}^w\}
\leq{\rm Pr}\{\bm{F}(\hat{P}_n^{Z^n})\in \Theta_\pm^l\}
\leq
|\mbb{Z}_{\Theta,\pm}^{w,l}|\cdot\max_{\mu\in\mbb{Z}_{\Theta,\pm}^{w,l}}  {\rm Pr}\{\bm{F}(\hat{P}_n^{Z^n})\in\mbb{R}_{\mu,l}^w\},
}
which implies
\alg{
-\lim_{n\rightarrow\infty}\frac{1}{n}\ln{{\rm Pr}\{\bm{F}(\hat{P}_n^{Z^n})\in \Theta_\pm^l\}}
=
-\lim_{n\rightarrow\infty}\frac{1}{n}\ln{\max_{\mu\in\mbb{Z}_{\Theta,\pm}^{w,l}}  {\rm Pr}\{\bm{F}(\hat{P}_n^{Z^n})\in\mbb{R}_{\mu,l}^w\}}.
}
Using \req{yakis}, the R.H.S.~of the above equality is further calculated to be
\alg{
\min_{\mu\in\mbb{Z}_{\Theta,\pm}^{w,l}}-\lim_{n\rightarrow\infty}\frac{1}{n}\ln{{\rm Pr}\{\bm{F}(\hat{P}_n^{Z^n})\in\mbb{R}_{\mu,l}^w\}}
=
\min_{\mu\in\mbb{Z}_{\Theta,\pm}^{w,l}}\inf_{\bm{t}\in\mbb{R}_{\mu,l}^w}\mfk{D}(\bm{t})
=
\inf_{\bm{t}\in\Theta_\pm^l}\mfk{D}(\bm{t}).
}
Thus, noting that $\Theta_-^l\subseteq\Theta\subseteq\Theta_+^l$, we have
\alg{
-\limsup_{n\rightarrow\infty}\frac{1}{n}\ln{{\rm Pr}\{\bm{F}(\hat{P}_n^{Z^n})\in \Theta\}}
\geq
-\limsup_{n\rightarrow\infty}\frac{1}{n}\ln{{\rm Pr}\{\bm{F}(\hat{P}_n^{Z^n})\in \Theta_+^l\}}
=
\inf_{\bm{t}\in\Theta_+^l}\mfk{D}(\bm{t})
}
and
\alg{
-\liminf_{n\rightarrow\infty}\frac{1}{n}\ln{{\rm Pr}\{\bm{F}(\hat{P}_n^{Z^n})\in \Theta\}}
\leq
-\liminf_{n\rightarrow\infty}\frac{1}{n}\ln{{\rm Pr}\{\bm{F}(\hat{P}_n^{Z^n})\in \Theta_-^l\}}
=\inf_{\bm{t}\in\Theta_-^l}\mfk{D}(\bm{t}).
}
From \req{eaeaso}, we have that
\alg{
\lim_{l\rightarrow\infty}\Theta_+^l
=
{\rm cl}\Theta,
\quad
\lim_{l\rightarrow\infty}\Theta_-^l
=
{\rm int}\Theta.
}
Since $\mfk{D}(t)$ is continuous by assumption, it follows that
\alg{
\lim_{l\rightarrow\infty}\inf_{t\in\Theta_+^l}\mfk{D}(t)
=
\inf_{t\in{\rm cl}\Theta}\mfk{D}(t),
\quad
\lim_{l\rightarrow\infty}\inf_{t\in\Theta_-^l}\mfk{D}(t)
=
\inf_{t\in{\rm int}\Theta}\mfk{D}(t).
}
All in all, we obtain \req{smetana1}, \req{smetana2} and complete the proof.
\QED

\section{Properties of the binary relative entropy }
\lapp{BRE}

We summarize the properties of the binary relative entropy.
For binary probability distributions $\{\alpha,1-\alpha\}$ and $\{\beta,1-\beta\}$, the binary relative entropy $\mathscr{D}$ is defined by 
\alg{
\mathscr{D}(\alpha\|\beta)
:=D(\{\alpha,1-\alpha\}\|\{\beta,1-\beta\})
=
\alpha\ln{\frac{\alpha}{\beta}}+(1-\alpha)\ln{\frac{1-\alpha}{1-\beta}}.
} 
It is nonnegative, and is equal to zero if and only if $\alpha=\beta$.
It is a convex function both in $\alpha$ and $\beta$.
When $\alpha\neq\beta$, it is finite if and only if $0<\beta<1$.
The derivatives of the binary relative entropy are calculated as
\alg{
\frac{\partial\mathscr{D}(\alpha\|\beta)}{\partial\alpha}
=\ln{\left(\frac{\alpha}{1-\alpha}\cdot\frac{1-\beta}{\beta}\right)},
\quad
\frac{\partial\mathscr{D}(\alpha\|\beta)}{\partial\beta}
=-\frac{\alpha}{\beta}+\frac{1-\alpha}{1-\beta}
\laeq{kiraku}
}
and
\alg{
\frac{\partial^2\mathscr{D}(\alpha\|\beta)}{\partial\alpha^2}
=\frac{1}{\alpha}+\frac{1}{1-\alpha},
\quad
\frac{\partial^2\mathscr{D}(\alpha\|\beta)}{\partial\beta^2}
=\frac{\alpha}{\beta^2}+\frac{1-\alpha}{(1-\beta)^2}.
}
It follows from \req{kiraku} that
\alg{
(\beta-\alpha)\frac{\partial\mathscr{D}(\alpha\|\beta)}{\partial\beta}
\geq0.
}
In particular, we have
\alg{
\left.\frac{\partial\mathscr{D}(\alpha\|\beta)}{\partial\alpha}\right|_{\alpha=\beta}
=
\left.\frac{\partial\mathscr{D}(\alpha\|\beta)}{\partial\beta}\right|_{\alpha=\beta}
=0
}
and
\alg{
\left.\frac{\partial^2\mathscr{D}(\alpha\|\beta)}{\partial\alpha^2}\right|_{\alpha=\beta}
=
\left.\frac{\partial^2\mathscr{D}(\alpha\|\beta)}{\partial\beta^2}\right|_{\alpha=\beta}
=\frac{1}{\alpha(1-\alpha)}.
}
Thus, in the second order of $\epsilon\ll1$, we have
\alg{
\mathscr{D}(\alpha\|\alpha+\epsilon)
\approx
\mathscr{D}(\alpha+\epsilon\|\alpha)
\approx
\frac{\epsilon^2}{2\alpha(1-\alpha)}.
}
%In particular, we have
%\alg{
%\min_\alpha\mathscr{D}(\alpha\|\alpha+\epsilon)
%\approx
%2\epsilon^2.
%}

\section{Proof of Eq.~\req{luzern} }
\lapp{prfnuw}

We prove that the solution of an equation
\alg{
\frac{1}{3}\sum_{k=-1,0,1}\frac{1}{1-w\nu+k\nu}=1
\laeq{yahay}
}
for $\nu$, under the condition $\nu\neq0$ and
\alg{
1-(w-k)\nu
\geq
0\;(k=-1,0,1),
\laeq{oosu}
}
is given by
\alg{
\nu^*(w)
=
\frac{1-3w^2-\sqrt{1+3w^2}}{3w(1-w^2)}.
}

From \req{yahay}, we have
\alg{
3w\nu^3-3w^3\nu^3+6w^2\nu^2-2\nu^2-3w\nu=0.
}
Noting that $\nu\neq0$, we obtain
\alg{
3w(1-w^2)\nu^2+2(3w^2-1)\nu-3w=0.
}
Thus, we have the following two candidates for the solution:
\alg{
\nu_{\pm}(w)
=
\frac{1-3w^2\pm\sqrt{1+3w^2}}{3w(1-w^2)}.
}
We prove that $\nu_-(w)$ is the only solution that satisfies the condition \req{oosu}, regardless of the value of $w$ satisfying $-1\leq w\leq1$.
To be explicit, we prove that
\alg{
1-(w+1)\nu_-(w)
&\geq
0,
\laeq{num1}
\\
1-w\nu_-(w)
&\geq
0,
\laeq{num2}
\\
1-(w-1)\nu_-(w)
&\geq
0
\laeq{num3}
}
for any $-1\leq w\leq1$
and
\alg{
1-(w+1)\nu_+(w)
&\leq
0\quad(w\geq0),
\laeq{nul1}
\\
1-(w-1)\nu_+(w)
&\leq
0\quad(w\leq0).
\laeq{nul2}
}
We will use the fact that
\alg{
&
|1-3w|
\leq
\sqrt{1+3w^2}
\leq
|1+3w|
\quad
(w\geq0),
\\
&
|1+3w|
\leq
\sqrt{1+3w^2}
\leq
|1-3w|
\quad
(w\leq0).
}
For $\nu_-(w)$, we have
\alg{
1-(w+1)\nu_-(w)
&=\frac{-(1-3w)+\sqrt{1+3w^2}}{3w(1-w)},
\\
1-w\nu_-(w)
&=\frac{2+\sqrt{1+3w^2}}{3(1-w^2)},
\\
1-(w-1)\nu_-(w)
&=\frac{(1+3w)-\sqrt{1+3w^2}}{3w(1+w)},
}
which yields \req{num1}-\req{num3}.
For $\nu_+(w)$, we have
\alg{
1-(w+1)\nu_+(w)
&=\frac{-(1-3w)-\sqrt{1+3w^2}}{3w(1-w)},
\\
1-(w-1)\nu_+(w)
&=\frac{(1+3w)+\sqrt{1+3w^2}}{3w(1+w)},
}
which yields \req{nul1} and \req{nul2}.
\QED

\section{Comparison with Previous Results }
\lapp{compPRPR}

We compare the exponents obtained for a few examples in the main sections to those obtained from the previous results based on Levy's lemma.
We compare the exponents for the maximum eigenvalue (Example 2 in \rSec{formresults}) and entropy (\rSec{RFentropy}) with those obtained from the results of \cite{hayden2006aspects}, and that for the expectation value (\rSec{expvalue}) with those deduced from the results of \cite{popescu2005foundations,reimann2015generalization}.
Note that the exp is taken base $2$ in \cite{hayden2006aspects}, while we adopt base $e$ here as in \cite{popescu2005foundations,reimann2015generalization}.

As for the maximum eigenvalue, Lemma III.4 in \cite{hayden2006aspects} asserts that for any $0<\varepsilon\leq1$, it holds that
\alg{
{\rm Pr}\left\{\left.\lambda_{\rm max}(\rho_\Psi)>\frac{1+\varepsilon}{m}\right|\Psi\sim\mu_n\right\}
\leq
\left(\frac{10m}{\varepsilon}\right)^{2m}\exp{\left(-\frac{n\varepsilon^2}{14}\right)}
\laeq{alwlg}
}
and
\alg{
{\rm Pr}\left\{\left.\lambda_{\rm min}(\rho_\Psi)<\frac{1-\varepsilon}{m}\right|\Psi\sim\mu_n\right\}
\leq
\left(\frac{10m}{\varepsilon}\right)^{2m}\exp{\left(-\frac{n\varepsilon^2}{14}\right)}.
} 
It is straightforward to verify that \req{alwlg} yields
\alg{
-\limsup_{n\rightarrow\infty}
\frac{1}{n}\ln{\rm Pr}\left\{\left.\lambda_{\rm max}(\rho_\Psi)>\frac{1+(m-1)\epsilon}{m}\right|\Psi\sim\mu_n\right\}
\geq
\frac{(m-1)^2\epsilon^2}{14}.
\laeq{arrarar}
}
As shown in Example 2 in \rSec{formresults}, the exact exponent is given by $m(m-1)\epsilon^2/2$.
This is an improvement over \req{arrarar} by factor $7/(1-1/m)$, when viewed as an upper bound on the probability of large deviation.

Theorem III.3 in \cite{hayden2006aspects} provides a large deviation bound for the entropy. It asserts that for any $3\leq m\leq n$, it holds that
\alg{
{\rm Pr}\left\{\left.S(\rho_\Psi)<\ln{m}-\frac{m}{n}-\epsilon\right|\Psi\sim\mu_n\right\}
\leq
\exp{\left(-\frac{(mn-1)\epsilon^2}{8\pi^2(\ln{m})^2}\right)}.
\laeq{taiky}
}
The equation \req{taiky} is equivalent to
\alg{
{\rm Pr}\left\{\left.1-\frac{S(\rho_\Psi)}{\ln{m}}>\frac{m}{n\ln{m}}+\varepsilon\right|\Psi\sim\mu_n\right\}
\leq
\exp{\left(-\frac{(mn-1)\varepsilon^2}{8\pi^2}\right)}.
}
Hence, for any $\delta>\kappa>0$ and sufficiently large $n$, it holds that
\alg{
{\rm Pr}\left\{\left.1-\frac{S(\rho_\Psi)}{\ln{m}}>\delta\right|\Psi\sim\mu_n\right\}
\leq
\exp{\left(-\frac{(mn-1)(\delta-\kappa)^2}{8\pi^2}\right)}.
}
It follows that
\alg{
-\limsup_{n\rightarrow\infty}\frac{1}{n}\ln{\rm Pr}\left\{\left.1-\frac{S(\rho_\Psi)}{\ln{m}}>\delta\right|\Psi\sim\mu_n\right\}
\geq
\frac{m(\delta-\kappa)^2}{8\pi^2}.
\laeq{laghlah}
}
Since this relation holds for any $\delta>0$, we arrive at
\alg{
-\limsup_{n\rightarrow\infty}\frac{1}{n}\ln{\rm Pr}\left\{\left.1-\frac{S(\rho_\Psi)}{\ln{m}}>\delta\right|\Psi\sim\mu_n\right\}
\geq
\frac{m\delta^2}{8\pi^2}.
}
We have shown in \rSec{RFentropy} that the exact exponent is, when $\delta$ is small, approximately equal to $\delta m\ln{m}$.
As an upper bound on the probability of large deviation, this is a quadratic improvement with respect to $\delta$ and an improvement by factor $8\pi^2\ln{m}$ over \req{laghlah}.

A large deviation bound for the expectation value can be deduced from Inequality (64) in \cite{popescu2005foundations}.
Let $\ca{H}_R\subseteq\ca{H}_S\otimes\ca{H}_E$ be a subspace with dimension $d_R$ and let $W$ be a Hermitian operator on $\ca{H}^S$. 
With $|\Psi\rangle\in\ca{H}_R$ randomly chosen with respect to the unitary invariant measure $\mu$ on $\ca{H}_R$, Ineq.~(64) in \cite{popescu2005foundations} reads
\alg{
{\rm Pr}\{\left.|{\rm Tr}[\rho_\Psi W]-{\rm Tr}[\pi W]|\geq w\right|\Psi\sim\mu\}
\leq2\exp\left(-\frac{d_Rw^2}{18\pi^3\|W\|_\infty^2}\right).
\laeq{umaku}
}
Let ${\rm dim}\ca{H}_S=m$, ${\rm dim}\ca{H}_E=n$ and $\ca{H}_R=\ca{H}_S\otimes\ca{H}_E$. 
We assume ${\rm Tr}[W]=0$ and $w>0$.
Then, \req{umaku} yields
\alg{
-\limsup_{n\rightarrow\infty}\frac{1}{n}\ln{\rm Pr}\{\left.{\rm Tr}[\rho_\Psi W]\geq w\right|\Psi\sim\mu_n\}
\geq
\frac{mw^2}{18\pi^3\|W\|_\infty^2}.\quad\quad
\laeq{murasa}
}
One can deduce a similar result from Equation (7) in \cite{reimann2015generalization}, in which case we have
\alg{
-\limsup_{n\rightarrow\infty}\frac{1}{n}\ln{\rm Pr}\{\left.{\rm Tr}[\rho_\Psi W]\geq w\right|\Psi\sim\mu_n\}
\geq
\frac{2mw^2}{9\pi^3\Delta_W^2},\quad\quad
\laeq{murasaki}
}
where $\Delta_W$ is the difference between the minimum and the maximum eigenvalues of $W$.
Noting that $\Delta_W\leq2\|W\|_\infty$, the bound \req{murasaki} is tighter than \req{murasa}.
We have shown in \rSec{expvalue} that the exact exponent is, when $w$ is small, approximately equal to $m^2w^2/2\|W\|_2^2$.
As an upper bound on the probability of large deviation, this is an improvement by factor $9\pi^3m\Delta_W^2/4\|W\|_2^2$ over \req{murasaki}. 
Note that we have $m\Delta_W^2\geq\|W\|_2^2$.

\bibliography{/Users/eyuriwakakuwa/Dropbox/DropTop/latexfiles/bibbib.bib}

 %Just because of unusual number of tables stacked at end
% Produces the bibliography via BibTeX.

\end{document}